\documentclass[letterpaper,twocolumn,10pt]{article}
\usepackage{usenix2019_v3}
\usepackage{tikz}
\usepackage{amsmath}
\usepackage{stfloats}
\usepackage{comment}
\usepackage{xspace}
\usepackage{filecontents}

\newcommand{\mypara}[1]{\vspace*{0.06in}\noindent\textbf{#1 }}

\usepackage{amsmath,amsfonts,bm,dsfont,cleveref}

\newcommand{\Ttwo}{T_2}
\newcommand{\Tone}{T_1}

\def\eqref#1{equation~\ref{#1}}

\def\1{\bm{1}}

\DeclareMathAlphabet{\mathsfit}{\encodingdefault}{\sfdefault}{m}{sl}
\SetMathAlphabet{\mathsfit}{bold}{\encodingdefault}{\sfdefault}{bx}{n}

\usepackage{booktabs}
\usepackage{multirow}
\usepackage{algorithm}
\usepackage{algpseudocode}
\usepackage{enumitem}
\usepackage{adjustbox}
\algdef{SE}[SUBALG]{Indent}{EndIndent}{}{\algorithmicend\ }%
\algtext*{Indent}
\algtext*{EndIndent}
\newcommand{\EiSL}{MIST\xspace}
\usepackage{colortbl}        
\usepackage{nopageno}

\begin{document}


\title{\Large \bf MIST: Defending Against Membership Inference Attacks Through Membership-Invariant Subspace Training}



\author{
{\rm Jiacheng Li}\\
Purdue University
\and
{\rm Ninghui Li}\\
Purdue University
 \and
 {\rm Bruno Ribeiro}\\
Purdue University
}

\maketitle

\begin{abstract}
In Member Inference (MI) attacks, the adversary tries to determine whether an instance is used to train a machine learning (ML) model.  MI attacks are a major privacy concern when using private data to train ML models.  Most MI attacks in the literature take advantage of the fact that ML models are trained to fit the training data well, and thus have very low loss on training instances.  Most defenses against MI attacks therefore try to prevent the model from heavily overfitting the training data.  Doing so, however, generally results in lower accuracy. 

We observe that training instances have different degrees of vulnerability to MI attacks.  Most instances will have low loss even when not included in training.  For these instances, they are less vulnerable to MI attacks and the model can fit them well without concerns.  An effective defense only needs to (possibly implicitly) identify instances that are vulnerable to MI attacks and avoids overfitting them.  A major challenge is how to achieve such an effect in an efficient training process. 

Leveraging two distinct recent advancements in representation learning: counterfactually-invariant representations and subspace learning methods, we introduce a novel Membership-Invariant Subspace Training (MIST) method to defend against MI attacks.  MIST avoids overfitting the vulnerable instances without significant impact on other instances.  We have conducted extensive experimental studies, comparing MIST  with various other state-of-the-art (SOTA) MI defenses against several SOTA MI attacks.  We find that MIST outperforms other defenses while resulting in minimal reduction in testing accuracy. 

\end{abstract}

\section{Introduction}

Neural network-based machine learning models are now prevalent in our daily lives, from voice assistants~\cite{lopez2018alexa}, to image generation~\cite{ramesh2021zero} and chatbots (e.g., ChatGPT-4~\cite{openai2023gpt4}). These large neural networks are powerful but also raise serious privacy concerns, such as whether personal data used to train these models are leaked by these models. 
One way to understand and address this privacy concern is to study membership inference (MI) attacks and defenses~\cite{shokri2017membership,nasr2019comprehensive}.  In MI attacks, an adversary seeks to infer if a given instance was part of the training data. 
If the optimal MI attack is merely as good as random guessing, then there is no privacy leakage in the sense of information theory~\cite{li2013membership}.

Most MI attacks in the literature are loss-based blackbox attacks, which exploit the fact that ML models tend to overfit the training data and have higher confidence on training instances. 
Several defenses were proposed to alleviate this effect by making the label of training instance ``softer'' so that the trained model would behave less confidently on training instances.  Shokri et al.~\cite{shokri2017membership} propose to use temperature in softmax function to increase the entropy of the predictions.  Li et al.~\cite{li2021membership} suggest to train with mix-up instances (linear combinations of two original training instances).  Chen and Pattabiraman~\cite{chen2023overconfidence} propose two methods to reduce confidence on training instances: (1) change labels of training instances to soft labels, and (2) use a regularizer to penalize low entropy predictions in training.  Forcing labels of training instances to be ``softer'', however, will inevitably incur some penalty of lower testing accuracy. 

We observe that these defenses fail to take advantage of the differences among training instances.  Some instances have low loss whether they are included in the training set or not, while other instances will incur high loss when they are not included in training~\cite{watson2021importance}. Hereafter, we will refer to these specific cases as ``distinctive''. 
Distinctive instances are naturally more vulnerable to MI attacks.  While this phenomenon has been exploited in MI attacks that set different loss thresholds for different instances~\cite{carlini2022membership,watson2021importance,wen2022canary}, existing defenses fail to take advantage of this effect.  By forcing all training instances to be ``softer'', these defenses incur accuracy loss that can be avoided. 

In this paper, we propose a novel training approach that naturally differentiates among instances with different degree of distinctiveness.  It can be viewed as making the label ``softer'' \textbf{only} for the distinctive instances, avoiding unnecessary penalty in testing accuracy.  Our approach uses ideas similar to subspace learning~\cite{jain2023dart}, and uses a novel regularizer to train models to be counterfactually invariant to the membership of each instance in the training data $D$. We thus call our approach Membership-Invariant Subspace Training (\EiSL).

The algorithm behind \EiSL is detailed in \Cref{sec:proposal} and here we give a brief description. \EiSL divides the training dataset into multiple subsets.  While one epoch of standard SGD can be viewed as sequentially updating the model on these subsets one by one, each epoch in \EiSL has three steps.  In the first step, we train on all subsets simultaneously, resulting in a subspace of multiple submodels (one for each subset). In the second step, for each such submodel $M_i$, we perform gradient updates to reduce the differences on instances used in computing $M_i$ between predictions given by $M_i$ and the average predictions given by other submodels.  In the third and last step, we average the resulting submodels.  The effect is that, if an instance used to train $M_i$ is not distinctive, $M_i$'s output on the instance would be similar to the average, resulting in little impact to the model.  On the other hand, if an instance is distinctive, the gradient update step would pull the model towards the average of other submodels, which approximates the counterfactual scenario where this instance is not used in training. 

%


\mypara{Contributions.} 

\begin{enumerate}[leftmargin=*] 
\item We propose the \textbf{Membership-Invariant Subspace Training} (\EiSL) defense designed to specifically defend against blackbox membership inference attacks on the most vulnerable instances in the training data (which also helps defending against overall blackbox MI attacks). \EiSL uses a novel two-phase subspace learning procedure that trains models regularized to be counterfactually invariant to the membership of each instance in the training data $D$.
\item We compare our defenses against baselines and provide extensive experiments showing that our defense can significantly reduce the effectiveness of existing membership inference attacks.
To the best of our knowledge, \EiSL is the first effective defense against LIRA and CANARY-style blackbox MI attacks that focuses on defending the most vulnerable instances in the training data, while improving the overall MI attack robustness.
\end{enumerate}

\mypara{Roadmap.} The rest of this paper is organized as follows. In section \ref{sec:prelim}, we introduce some machine learning background and summarize existing membership inference attacks and defenses in centralized setting. In section \ref{sec:defense}, we propose our new defense, the \textbf{Membership-Invariant Subspace Training} (\EiSL) defense. In section \ref{sec:experiments}, we present our extensive experimental evaluations and compare our proposed defense with existing defenses. 
In section \ref{sec:conclusion}, we summarize the findings presented in this paper. In section \ref{sec:future}, we discuss limitations of the proposed MIST defense and potential future directions.

\section{Preliminary}
\label{sec:prelim}


In this section, we describe the MI attacks that we use in our experimental evaluation and defense mechanisms that we compare against.  We focus on blackbox attacks, where the adversary queries the target model and uses the model prediction to infer membership.  Other works in this area are discussed in Section~\ref{sec:related} in the Appendix.

\subsection{MI attacks in blackbox setting}
\label{sec:mi_attacks_summary}

We classify these attacks based on what they use to query the target model $F^T$. 

\subsubsection{Using the Target Instance}

In these attacks, one uses the target instance $x$ to query the model.


\mypara{LOSS} (Using loss with a global threshold). 
Yeom et al.~\cite{yeom2018privacy} introduced an attack that predicts $x$ is a member when the target model's loss on $x$ is below the average training loss. 

\mypara{Class-NN} (Training class-specific Neural Networks for MI).  
Shokri's attack~\cite{shokri2017membership} trains multiple neural network-based membership classifiers, one for each class.  Training data are obtained using the shadow-mode technique, which is widely used in later attacks. 

Using the \textbf{shadow-model} technique, one assumes that the adversary knows a dataset $D^A$, which contains the target instance and is from the same distribution as the dataset used to train $F^T$. The adversary creates $k$ subsets $D_1,D_2,\ldots,D_k$ from $D^A$, and uses the same process used for training $F^T$ to train $k$ models, one from each $D_i$.  These are called shadow models. For each instance $x$, some shadow models were trained using $x$, and others were trained without.  The predictions of these models on instances in $D^A$ provide training data for membership classifiers. 

\mypara{Modified Entropy.} 
Song et al.~\cite{song2019membership} proposed to use a class-specific threshold on a modified entropy measure based on the model prediction to determine membership.  This can be viewed as a simplified version of the attack in~\cite{shokri2017membership}, and is very similar to using a class-specific loss threshold to determine membership. 


\mypara{LIRA.} Carlini et al.~\cite{carlini2022membership} proposed an instance-specific threshold attack.
For each instance, from the shadow models, one obtains a distribution for losses from models trained with $x$, and another distribution from models trained without $x$.  A threshold can be chosen using likelihood ratio.  Carlini et al.~suggest to choose the threshold that optimizes for attack effectiveness at a very low false positive rate.  We refer this attack as LIRA attack in later sections.

\subsubsection{Using Perturbations of the Target Instance}


\mypara{Random perturbations.} Jayaraman et al. ~\cite{jayaraman2021revisiting} proposes an attack that generates multiple perturbed instances by adding Gaussian noise to $x$, and then query $F^T$ using these perturbed instances and count how many times the prediction loss of them is higher than that of $x$. The instance $x$ is predicted to be a member if the count is beyond a threshold. We refer this attack as the random-perturbation attack in later sections.

\mypara{CANARY.} The attack proposed in Wen et al.~\cite{wen2022canary} also uses shadow models.  For each target instance $x$, the shadow models are partitioned into two sets: those trained with $x$, and those trained without.  One then computes a set of canaries, each generated via gradient descent starting from a slightly perturbed version of $x$, searching for an $x'$ such that the difference between the average losses of $x'$ from models in the two sets are as large as possible.  One then use these canaries to query the target model and use the loss to predict membership. We call this attack the CANARY attack in later sections.


\mypara{Adversarial perturbation label only.}
Choquette-Choo et al.~\cite{choquette2021label} proposed two attacks for the situations where only the predicted label is provided and identified the adversarial perturbation attack as more effective.  In this attack, one applies the adversarial example generation technique from~\cite{carlini2017towards} to generate $x'$ that is close to $x$ while have a different predicted label by $F^T$.  The attack predicts membership if $||x'-x||_2$ is high. 



\subsection{Defenses against MI Attacks}
\label{sec:mi_defense_summary}

\mypara{Adversarial Regularization (Adv-reg).}
Nasr et al.~\cite{nasr2018machine} proposed 
a defense that uses similar ideas as GAN.  
The classifier is trained in conjunction with an MI attacker model.  
The optimization objective of the target classifier is to reduce the prediction loss while minimizing the MI attack accuracy. 

\mypara{Mixup+MMD.}
Li et al.~\cite{li2021membership} proposed a defense that combines mixup data augmentation and MMD (Maximum Mean Discrepancy~\cite{fortet1953convergence,gretton2012kernel}) based regularization.  Instead of training with original instances, mixup data augmentation uses linear combinations of two original instances to train the model.  It was shown in~\cite{zhang2017mixup} that this can improve target model's generalization.   Li et al.~\cite{li2021membership} found that they also help to defend against MI attacks. Li et al.~\cite{li2021membership} also proposes to add a regularizer that is the MMD between the loss distribution of members and the loss distribution of a validation set not used in training.  This helps make the loss distribution of members to be more similar to the loss distribution on non-members.  

\mypara{Distillation for membership privacy (DMP).}
Distillation uses labels generated by a teacher model to train a student model.  It was proposed in~\cite{hinton2015distilling} for the purpose of model compression. 
Shejwalkar et al.~\cite{shejwalkar2021membership} proposed to use distillation to defend against MI attacks. One first trains a teacher model using the private training set, and then trains a student model using another unlabeled dataset from the same distribution as the private set. The intuition is that since the student model is not directly optimized over the private set, their membership may be protected. the authors also suggested to train a GAN using the private training set and draw samples from the trained GAN to train the student model, when no auxiliary unlabeled data is available. 

\mypara{SELENA.} Tang et al.~\cite{tang2022mitigating} proposed a framework named SELENA. 
One first generates multiple (overlapping) subsets from the training data, then trains one model from each subset.  One then generates a new label for each training instance, using the average of predictions generated by models trained without using that instance.  Finally, one trains a model using the training dataset using these new labels. 


\mypara{HAMP.} Chen et al.~\cite{chen2023overconfidence} proposed a defense combining several ideas. First, labels for training instances are made smoother, by changing $1$ to $\lambda$ and each $0$ to $\frac{1-\lambda}{k-1}$, where $\lambda$ is a hyperparameter and $k$ is the number of classes.  Second, an entropy based regularizer is added in the optimization objective.  
Third, the model does not directly return its output on a queried instance $x$.  Instead, one randomly generates another instance, reshuffle the prediction vector of the randomly generated instance based on the order of the probabilities of $x$ and returns the reshuffled prediction vector.  In essence, this last defense means returning only the order the classes in the prediction vector but not the actual values. 

\mypara{Mem-guard.}  Jia et al.~\cite{jia2019Memguard} proposed the \textbf{Mem-guard} defense.  In this defense, one trains an MI attack model in addition to the target classifier.  When the target classifier is queried with an instance, the resulting prediction vector is not directly returned.  Instead, one tries to find a perturbed version of the vector such that the perturbation is minimal and does not change the predicted label, and the MI attack model output (0.5,0.5) as its prediction vector.  

\mypara{Differential Privacy.}
Differential privacy (DP)~\cite{dwork2008differential,dwork2006calibrating,Dwo06} is a widely used privacy-preserving technique.  DP based defense techniques, such as DP-SGD~\cite{abadi2016deep}, add noise to the training process. This provides a theoretical upper-bound on the effectiveness of any MI attack against any instance. Unfortunately, achieving a meaningful theoretical guarantee (e.g., with a resulting $\epsilon < 5$) requires the usage of very large noises.  However, model trainer could use much smaller noises in DP-SGD.  While doing this fails to provide a meaningful theoretical guarantee (the $\epsilon$ value would be too large), this can nonetheless provide empirical defense against MI attacks. In~\cite{li2021membership,tang2022mitigating,chen2023overconfidence}, extensive experiments have shown that several other defenses can provide better empirical privacy-utility tradeoff than DP-SGD.




\section{Proposed Defense: Membership-Invariant Subspace Training}
\label{sec:defense}

Almost all MI attacks take advantage of the fact that training aims to reduce the loss on member instances to zero.  Many defenses thus try to deviate from this optimization objective.  Unfortunately, this usually results in significant testing accuracy drop.  
Our key insight is that we can take advantage of the fact that relatively few number of instances are truly vulnerable to MI attacks.  For example, under the LIRA attack at 0.001 False Positive Rate, less than 5\% of members are identified.  That is, for most instances, whether it is included in the training set or not will not drastically affect their loss.  Intuitively, we 
only need to change the optimization objective for the vulnerable instances to avoid overfitting them.  This avoids suffering from unnecessary testing accuracy reduction.  The challenge is how to achieve this effect in the training process without incurring a very high runtime overhead. 


\subsection{Threat Model}

We consider a blackbox attacker, who can query the target model and get its prediction, but does not use the internal state of the target model, such as model parameters. We also assume that the adversary knows the model architecture, the training recipe, and the distribution of training data, and is capable of training shadow models. This type of adversary is able to get the member prediction distribution and nonmember prediction distribution for a particular instance of interest by training multiple shadow models and then perform the LIRA attack or the CANARY attack, which are the strongest blackbox MI attacks in the literature. 


\subsection{Our Proposal}
\label{sec:proposal}


To effectively protect the membership of vulnerable instances without incurring accuracy reduction for less vulnerable instances, our proposed defense capitalizes on the combined strengths of two distinct recent advancements in representation learning: Counterfactually-invariant representations~\cite{mouli2022asymmetry,quinzan2022learning,veitch2021counterfactual,jiang2022invariant} and subspace learning methods~\cite{wortsman2021learning,jain2023dart}.

A {\em counterfactual dataset} is a set of data samples whose distribution has been changed by an intervention.
Inspired by differential privacy, given a training dataset $D$, we define $n=|D|$ environments $D_1,D_2,\cdots,D_n$ such that 
$D_M = D \setminus \{(x_M,y_M)\}$ for $1\leq M\leq n$. That is, each $D_M$ has the same distribution as the training data $p(x,y)$, except for the intervention that the $M$-th training example $(x_M,y_M)$ has probability zero.  

Let $F(\cdot; \theta)$ denote a classifier with parameters $\theta$ such that $F(x ;\theta)$ gives a vector of predicted probabilities for the classes.  Let $T_F(\cdot)$ denote the training algorithm such that $\theta_D=T_F(D)$ is the parameters trained using dataset $D$.  Ideally, we want $T_F$ to be counterfactually-invariant to the interventions, i.e., $\forall M \in \{1,\ldots,n\}$,
$F(x;T_F(D))$ and $F(x;T_F(D_M)$ have the same classification behavior. 
Note that this objective is related to yet different from the objective of $\epsilon$-differential privacy (DP)~\cite{dwork2008differential,dwork2006calibrating,Dwo06}. 
This objective is less strict than $\epsilon$-DP when $\epsilon=0$, which requires $T_F(D)$ and $T_F(D_M)$ to have exactly the same distribution.  Here we expect that $\theta_D=T_F(D)$ and $\theta_{D_M} = T_F(D_M)$ to have the same classification behavior, even though the parameters may be different. 



We will denote our type of counterfactual invariance as {\em membership-invariance}.
Employing membership-invariant classifiers as a defense against MI attacks appears both promising and ``straightforward'' at first glance. 
Closer inspection, however, reveals that the task of training a membership-invariant classifier (that generalizes well in test) to be immensely challenging, especially over large training datasets. 
Standard invariant representation training methods 
include invariant risk minimization~\cite{irm}, adversarial training, such as Ganin et al.~\cite{ganin2016domain}, that uses a minimax game to ensure representations cannot predict the environments, and the Hilbert-Schmidt conditional independence criterion~\cite{park2020measure} used by Quinzan et al.~\cite{quinzan2022learning} for counterfactually-invariant predictors).  Using these methods for training membership-invariant classifiers would require constructing $n$ leave-one-out datasets from $D$, each a new environment.  This is infeasible for large datasets commonly used in state-of-the-art models. 


The main technical challenge we need to overcome is thus: {\em How can we efficiently learn membership-invariant classifiers to defend against MI for all vulnerable instances?}


\subsubsection{Cross-difference loss}
\label{sec:cross}
We start by defining a regularization that we call {\em cross-difference loss} as follows:
\begin{equation}
\begin{aligned}
\label{eq:cross_diff_sup}
    \text{xdiff}&(\theta_D,\theta_{D\setminus \{(x_M,y_M)\}}) \\
& =\sup_{(x,y) \in \Omega} \Vert F(x;\theta_D)_y - F(x; \theta_{D \setminus \{(x_M,y_M)\}})_y \Vert_1,
\end{aligned}
\end{equation}
where $\Omega$ is the support of $p(x,y)$ and $\Vert \cdot \Vert_1$ is the $L_1$ norm.
\Cref{eq:cross_diff_sup} pushes the classifier $F(\cdot;\theta_D)$ trained with $(x_M,y_M)$ to have the same output as the classifier $F(\cdot;\theta_{D\setminus \{(x_M,y_M)\}})$ trained without $(x_M,y_M)$.
%
Then, any classifier $F$ that satisfies 
\begin{equation}
\label{eq:envinvmax}    
\sum_{M \in \{1,\ldots,n\}} \text{xdiff}(\theta_{D},\theta_{D \setminus \{(x_M,y_M)\}}) = 0,
\end{equation}
{\em is membership-invariant}.
This is easy to verify since the condition in \Cref{eq:envinvmax} implies $F(x; \theta_{D \setminus \{(x_M,y_M)\}}) = F(x; \theta_{D}) = F(x; \theta_{D \setminus \{(x_{M'},y_{M'})\}})$, for all $M,M' \in \{1,\ldots,n\}$, that is, the classifier output is invariant to the environment in which it was trained.

\subsubsection{Membership-invariance via Two-Phase Subspace Learning}
\label{sec:vfl}
The challenge of training a classifier that satisfies \Cref{eq:envinvmax} is the computational cost.
To support the fast optimization of $F$ using the cross-difference loss in \Cref{eq:envinvmax}, we will make use of the concept of subspace learning~\cite{wortsman2021learning}.
Subspace learning optimizes neuron weights in a subspace that contains
diverse solutions that can be ensembled, approaching the ensemble performance of {\em independently} trained neural networks without the associated training cost.
Originally designed by Wortsman et al.~\cite{wortsman2021learning} to boost accuracy, calibration, and robustness to label noise, we will repurpose  subspace learning for learning membership-invariant classifiers.
In particular, our approach (\EiSL) leverages a version of subspace learning based on (virtual) federated learning: DART (Diversify-Aggregate-Repeat Training)~\cite{jain2023dart}.

Our core idea adapts DART and simultaneously trains multiple diverse models that are regularized with a computationally efficient approximation of the regularization implied by \Cref{eq:envinvmax}. These diverse models within the subspace, in effect, regularize each other. They impose penalties on each other if discrepancies arise in their outputs when trained on their respective subsets of the training data. We denote the resulting method \EiSL and detail it in \Cref{alg:vfl}; we also summarize the method below.

\mypara{\EiSL.}  Training in \EiSL consists of $E$ global epochs.  We use 
$\overline{\theta}_e$ to denote (parameters of) the aggregated (i.e., global) model at epoch $e = 0,\ldots, E$. 
$\overline{\theta}_0$ is initialized with random parameters. 
The $e$-th epoch (where $1\le e\le E$) consists of the following steps.

\textbf{Initialization: Data Partitioning.} The training data $D$ is split into $C$ disjoint subsets $D^c$, $c =1 ,\ldots, C$.  We use $\overline{\theta}_{e-1}$ to initialize $C$ (local) models. 

\textbf{Phase 1: Exploring diverse models (local training).} 
For each $c \in \{1,\ldots,C\}$, we use $D^c$ to update the $c$-th local model, by performing gradient updates $\Tone \geq 1$ times, where $\Tone$ is a hyperparamenter. 
We use $\theta^c_e$ to denote (parameters of) the $c$-th local model after the gradient updates.

\textbf{Phase 2: Finding models with small cross difference loss.} 
In Phase 2, we aim to find models near the diverse models found in Phase 1 that have low cross-difference loss. 
For $c=1,\ldots,C$, let $w^c_0=\theta^c_e$, we perform $T_2$ gradient update steps, using the following approximation of the cross-difference loss (described in \Cref{eq:cross_diff_sup}) to regularize $w^c_t$ in \Cref{alg:vfl}:
 \begin{equation}
     \label{eq:cross_diff}
     \text{xdiff}_c(w_{t}^c)\! = \!\!\!\!\! \sum_{(x,y) \in D^c} \bigg\Vert F(x;w_{t}^c)_y - \frac{\sum_{i \neq c}F_{}(x; \theta_{e}^i)_y}{C-1}\bigg\Vert_{1}.
 \end{equation}
Minimizing \Cref{eq:cross_diff} 
will push our models towards satisfying \Cref{eq:envinvmax}.
Note that if an instance $(x,y)\in D^c$ is not vulnerable to MI attacks, the confidence from the $c$-th local model (trained using $(x,y)$) $F(x;w_{t}^c)_y$ will be similar to the average confidence from other local models, and 
the $L_1$ norm in \Cref{eq:envinvmax} is close to $0$, resulting in little change to the model parameter.  If, however,  $(x,y)\in D^c$ is vulnerable to MI attacks, then the $L_1$ norm in \Cref{eq:envinvmax} will be large, resulting the $c$-th local model to be updated to fit less well on $(x,y)$.


In \Cref{eq:cross_diff} we are not computing the $L_1$ loss over the entire support $\Omega$ of the training distribution $p(x,y)$ as in \Cref{eq:cross_diff_sup} but, rather, restricting the cross-difference loss to the examples in $D^c$. This is a reasonable approximation since these examples are the ones used to directly optimize the $c$-th local model, and we thus focus on protecting their membership.

In \Cref{eq:cross_diff}, we use $L_1$ norm.  In Section \ref{sec:ablation}, we compare the effect of using $L_1$, $L_2$ norms, and KL divergence in an ablation study.

\textbf{Model aggregation.} The (global) training epoch $e \geq 1$ ends by averaging the parameters of all $C$ local models into a new  parameter $\overline{\theta}_{e}$.  In general, as in subspace learning, the averaging can be a convex combination of the model parameters.  In our experiments we just use the arithmetic mean $\overline{\theta}_{e} = (1/C) \sum_{c=1}^C \theta_{e}^c$. 

\mypara{Returning.} At the end of our \EiSL algorithm, we return $\overline{\theta_E}$ as the final trained model parameters.
%

\begin{algorithm}
\caption{(\EiSL) Membership-Invariant Subspace Training}\label{alg:vfl}
\begin{algorithmic}
\State \textbf{Input}: $C$: the number of local models; $D$: the training dataset; $E$: the number of epochs.  Hyperparameters $\Tone$, $\Ttwo$, and $\lambda$ (weight for cross difference loss). 

\vspace{0.2cm}
\State Initialize $\overline{\theta}_0$ \Comment{Model initialization}
\State \textbf{for $e = 1$ to $E$ do} 
\Indent
\State Partition $D$ into $D^c$, where $1 \leq c \leq C$
\State \textbf{for $c = 1$ to $C$ do}
\Comment{can be performed in parallel}
\Indent 
\State \text{$\theta^c_e=\text{local\_training}(\overline{\theta}_{e-1}$,$D^{c}$,$\Tone$)}\
\EndIndent
\State \textbf{for $c = 1$ to $C$ do}
\Comment{can be performed in parallel}
\Indent 
\State \text{$\theta^c_e=\text{xdifference\_update}(c$,${\theta}^{1}$$,\ldots,$${\theta}^{C}$,$D^{c}$,$\lambda$,$\Ttwo$)}
\EndIndent
\State  $\overline{\theta}_e = \frac{\sum_{c=1}^C \theta^c_e}{C}$ \Comment{Aggregate and average model weights}
\EndIndent
\State \textbf{Output:} $\overline{\theta_E}$
\vspace{0.2cm}

\State \textbf{Function} local\_training($\overline{\theta}$,$D^{c}$,$\Tone$):
\Comment{Phase 1}
\Indent
\State $w^c_{0} = \overline{\theta}$
\State \textbf{for $t = 1$ to $\Tone$ do}
\Indent
\State $\text{gradient updates to $w_{t}^c$ to minimize } \frac{1}{|D^{c}|}  \mathcal{L}(w_{t-1}^c,D^c)$
\EndIndent
\EndIndent
\textbf{Output:} $w_{\Tone}^c$

\State \textbf{Function} xdifference\_update($c$, ${\theta}^{1}$$,\ldots,$${\theta}^{C}$,$D^{c}$,$\lambda$,$\Ttwo$):
\Comment{Phase 2}
\Indent
\State $w^c_{0} = \theta^c$
\State \textbf{for $t = 1$ to $\Ttwo$ do}
\Indent

\State $\text{gradient updates to $w_{t}^c$ to minimize } \frac{\lambda}{|D^{c}|} \, \text{xdiff}_c(w_{t-1}^c)$
\EndIndent
\EndIndent
\textbf{Output:} $w_{\Ttwo}^c$

\end{algorithmic}
\end{algorithm}

\subsubsection{Further improvements}
\EiSL can be augmented with some existing MI defense approaches.
For instance, the Mix-up data augmentation MI defense  proposed by Li et al.~\cite{li2021membership} can be easily integrated into \EiSL (\Cref{alg:vfl}). 
Li et al.~\cite{li2021membership} shows that Mix-up data augmentation can reduce the generalization gap between train and test data, consequently making black-box MI attacks more difficult. 
Mix-up training uses linear interpolation of two different training instances to generate a mixed instance and train the classifier with the mixed instance. The generation of mixed instances can be described as follows:
        \begin{align}
        \tilde{x} & = \beta x_i + (1-\beta) x_j, \nonumber\\ 
        \tilde{y} & = \beta y_i + (1-\beta) y_j, \label{eq:mixup}
    \end{align}
    where $\beta \sim \text{Beta}(\alpha,\alpha), \alpha \in (0,\infty).$
    Here $x_i$ and $x_j$ in \Cref{eq:mixup} are instance feature vectors randomly drawn from the training set; $y_i$ and $y_j$ are one-hot label encodings corresponding to $x_i$ and $x_j$. The new tuple $(\tilde{x},\tilde{y})$ is used in training. 


\section{Evaluation}
\label{sec:experiments}
In this section we present our extensive experimental results. We start by describing the details of our experimental setup. Next we discuss the computational cost of considered defenses. Then we dive into the detailed comparison between MIST and other defenses. After this, we present the evaluation of MIST against label-only attacks. In the following subsection, we show how to choose hyperparameters for our MIST defense. Lastly, we present ablation studies.

\subsection{Experimental Setup} \label{sec:exp_setting}

\subsubsection{Datasets}
Our evaluation uses the datasets most commonly used in MI attack literature~\cite{shokri2017membership,nasr2018machine,chen2023overconfidence,salem2018ml,nasr2019comprehensive,yeom2018privacy,jia2019Memguard}. 

\mypara{CIFAR-10 and CIFAR-100.} They contain 60,000 color images of size 32 $\times$ 32, divided into 50,000 for training and 10,000 for testing.  In CIFAR-100, these images are divided into 100 classes, with 600 images for each class.  In CIFAR-10, these 100 classes are grouped into 10 more coarse-grained classes; there are thus 6000 images for each class. We use AlexNet, ResNet-18 and DenseNet-BC(100,12) (abbreviated as DenseNet), which are the standard neural networks used in prior work~\cite{nasr2019comprehensive,li2021membership}. The model architectures and hyperparameters for training are described in \Cref{tab:training_recipe} in the Appendix. 
    
\mypara{TEXAS-100.} The records in the dataset contain information about inpatient stays in several health care facilities published by the TEXAS Department of State Health Services. 
We obtained the dataset from the authors of~\cite{shokri2017membership}.  The dataset contains 67,330 records and 6,170 binary features.  The records are clustered into 100 classes, each representing a different type of patient.   We use the fully connected neural network from~\cite{nasr2019comprehensive} as the target model.

\mypara{PURCHASE-100.} This dataset is based on the ``acquire valued shopper'' challenge from Kaggle. This dataset includes shopping records for several thousand individuals.  We obtained the processed and simplified version of this dataset from the authors of~\cite{shokri2017membership}.  Each data instance has 600 binary features. This dataset is clustered into 100 classes and the task is to predict the class for each customer. The dataset contains 197,324 data instances. We use the fully connected neural network from~\cite{nasr2019comprehensive} as the target model.

\mypara{LOCATION.} This dataset contains the location “check-in” records of different people. It has 5,010 data records with 446
binary features, each of which corresponds to a certain location type and indicates whether the individual has visited that
particular location. The goal is to predict the user’s geo-social type. There are 30 classes in this dataset.  We use the fully connected neural network from~\cite{nasr2019comprehensive} as the target model.

\begin{figure*}
    \centering
    \includegraphics[height=19.5cm,width=0.9\textwidth]{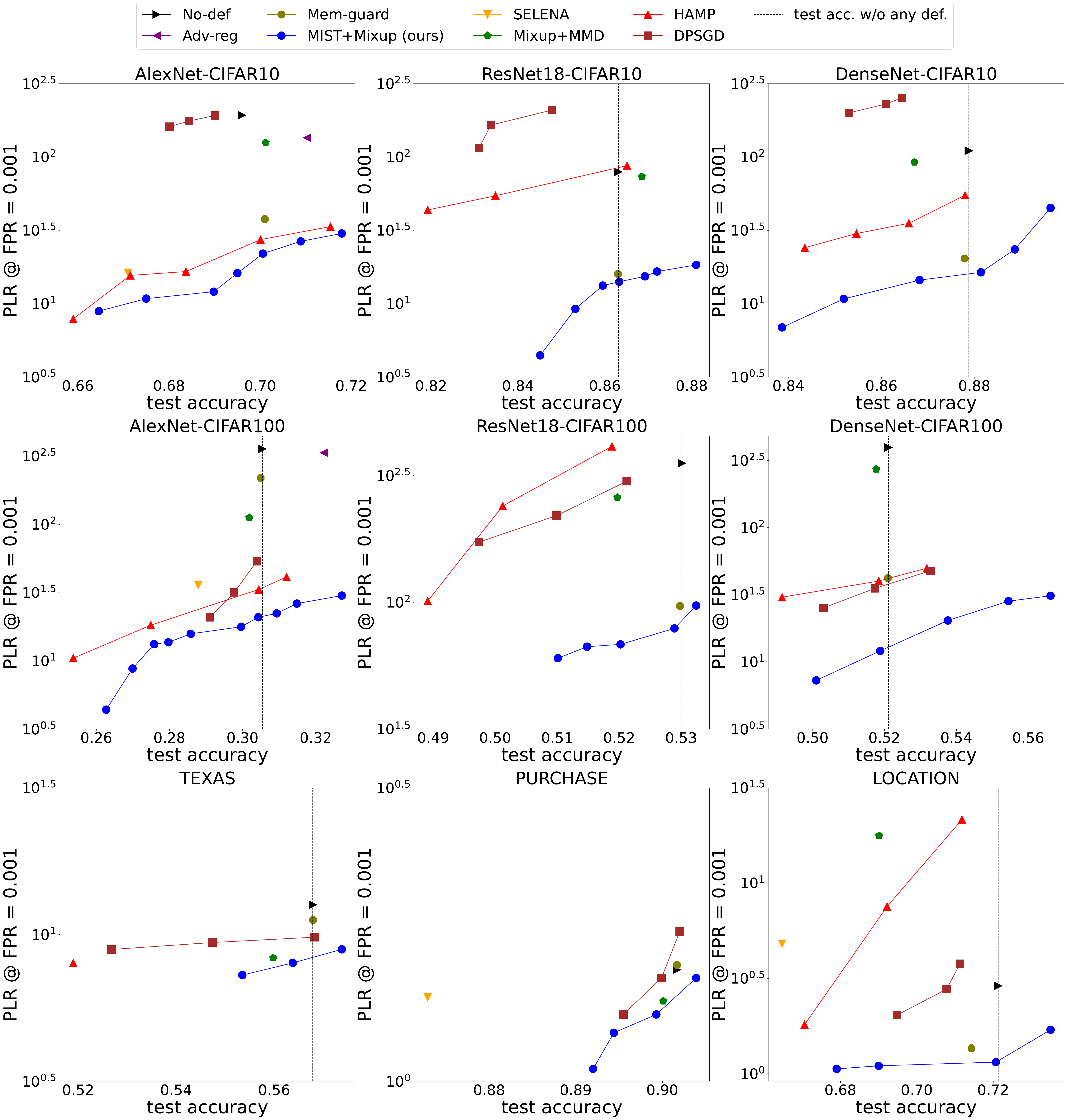}
    \caption{Comparing all defenses using the highest MI attack PLR @ 0.001 FPR among all evaluated attacks. {\bf Defenses placed at the lower right corner are better} (high test accuracy and low PLR). Notice that for PURCHASE, TEXAS and LOCATION datasets the mixup data augmentation is not applied. We exclude some results when the test accuracy drop is larger than $5\%$ for clearer comparison.}
    \label{fig:all_def_comp}
\end{figure*}

\subsubsection{Evaluated Attacks and defenses}


\mypara{Attacks.} We consider the six attacks described in Section~\ref{sec:mi_attacks_summary}: LOSS~\cite{yeom2018privacy}, Modified Entropy~\cite{song2019membership}, Class-NN~\cite{shokri2017membership}, random perturbation attack from~\cite{jayaraman2021revisiting}, the online version of the LIRA attack from~\cite{carlini2022membership} and the CANARY attack from ~\cite{wen2022canary}. Readers may want to review Section~\ref{sec:mi_attacks_summary} on how these attacks work. The latter two are the state-of-the-art blackbox membership inference attack. We set the number of shadow models to be $100$, which is common in the literature. We assume that the attackers are adaptive attackers, which means that the attackers already know MIST defense mechanism (including hyperparameters) and can train similar models in exactly the same setting.

\mypara{Defenses.} We compare with the following existing defenses: adversarial regularization (adv-reg)~\cite{nasr2018machine}, Mem-guard~\cite{jia2019Memguard}, DMP~\cite{shejwalkar2021membership}, Mixup+MMD~\cite{li2021membership}, SELENA~\cite{tang2022mitigating}, HAMP~\cite{chen2023overconfidence} and DP-SGD~\cite{abadi2016deep}. Readers may want to review Section~\ref{sec:mi_defense_summary} on how these defense methods work. As for our MIST defense, we use the following setting: for $\Tone$, we set it to be the number so that each data instance in subset $D^c$ can be visited just once and we set $\Ttwo = \Tone$. The chosen $\lambda$ for each model and dataset is also listed in \Cref{tab:training_recipe}. We first tune the $C$ to achieve the highest validation accuracy possible then we tune the $\lambda$ so that the test accuracy drop is less than $1\%$.

\subsubsection{Evaluation Metrics}
\label{sec:eval_metrics}
Following existing literature~\cite{nasr2019comprehensive,carlini2022membership}, we use a balanced evaluation set. For CIFAR-10 and CIFAR-100 datasets, we assume that the model trainer has 20,000 data instances to perform the model training. For PURCHASE dataset and TEXAS dataset, we assume the model trainer has 40,000 training data instances. For LOCATION dataset, we assume the model trainer has 4,000 training data. We mainly use PLR (positive likelihood ratio, which is the ratio between TPR and FPR where lower means a more effective defense) at a fixed low FPR as suggested in Carlini et al.~\cite{carlini2022membership} to evaluate different attacks. For completeness, we also include AUC (area under curve) scores in the Appendix. 


\subsection{The computational cost (time and memory) of defenses}
Here we assume that all the hyperparameters are already chosen, and compare the time that that model optimization needs to produce a final usable model and predictions for the test set. We run this comparison experiments on a server with one GTX-1080Ti GPU and we use CIFAR-100 on AlexNet as an example to compare. \Cref{tab:run_time_comp} shows our results, where we can see that MIST achieves the second fastest running time among all defenses (only slower than HAMP defense). Note that the SELENA defense requires significantly more time and space comparing to the other defenses, since it needs to train many extra models (25 with the code's default parameters) and store them. Moreover, the Mem-guard defense also requires significant time, because the noise for each prediction needs to be optimized separately. As for memory cost of our MIST defense, it depends on the number of models $C$ (usually $2$ to $4$).  The training time of MIST is about twice that of the baseline because at each epoch MIST performs two backprop steps instead of one to perform the regularization step.
In conclusion, MIST requires moderate extra time and memory for model training where the best privacy-utility tradeoff is provided. 

\begin{table}[]
    \centering
    \resizebox{.95\columnwidth}{!}{
    \begin{tabular}{lrr}
        \toprule
        Defense name & Running time (s) & memory cost factor \\
        \midrule
        SGD baseline & 1195 & 1 \\
        Adv-reg & 7008 & 1 \\
        DMP & 2543 & 2 \\
        Mem-guard & 36373 & 1 \\
        SELENA & 20320 & 26 \\
        Mixup+MMD & 3154 & 1 \\
        HAMP & 1280 & 1  \\
        MIST+Mixup & 2139 & 4 \\
        \bottomrule
    \end{tabular}
    }
    \caption{Computational cost comparison for different defenses. For memory cost, we only count the number of models $C$ to be trained.}
    \label{tab:run_time_comp}
\end{table}

\subsection{Defense effectiveness evaluation}


\Cref{fig:all_def_comp} presents the experimental results comparing MIST with other defenses on 5 datasets (with three model architectures for each of CIFAR-10 and CIFAR-100).  We include only data points where the test accuracy drop is no more than than $5\%$ for clearer comparison. 
In \Cref{fig:all_def_comp}, the $x$-axis gives test accuracy and the $y$-axis gives PLR$_{0.001}$ (PLR when FPR$=0.001$) in log-scale.  We prefer methods to be in the lower-right region, as they can provide higher test accuracy under the same PLR$_{0.001}$.  Note that our proposed defense method MIST+Mixup performs the best in all experiments. 
All the detailed PLR numbers are presented in \Cref{tab:full_def_eval_tpr_cifar10,tab:full_def_eval_tpr_cifar100,tab:full_def_eval_tpr_binary}. 

When PLR$_{0.001}=100$, it means that when we ensure that no more than 0.1\% of non-members are falsely identified as members, if we identify 101 instances as members, we can expect that 100 are indeed members, and 1 of them is a false positive.  When PLR$_{0.001}=2$, it means that when we ensure that no more than 0.1\% of non-members are falsely identified as members, if we identify 3 as members, we can expect 2 of them are indeed members. 

\mypara{Effect of MIST+Mixup.}
We first look at how well our proposed defense (MIST+Mixup) prevents MI attacks. 
For CIFAR-10, our defense can reduce the highest PLR among all attacks from $127.63$ to $17.48$ (averaged over all models); for CIFAR-100 dataset, our defense can reduce the highest PLR from $368.94$ to $29.97$ (averaged over all models). For PURCHASE dataset, TEXAS dataset and LOCATION dataset, the highest PLR is reduced from $6.51$ to $4.10$, $1.55$ to $1.08$, $2.89$ to $1.15$ respectively. 
As for the average case attack effectiveness, our defense can also provide noticeable AUC drop comparing to the case without any defense, which is detailed in  \Cref{tab:full_def_eval_auc_cifar10,tab:full_def_eval_auc_cifar100,tab:full_def_eval_auc_binary} in the Appendix. In summary, our new defense can provide significant attack effectiveness (including PLR and AUC) reduction for highly vulnerable cases (CIFAR-10 and CIFAR-100) and provide further attack effectiveness reduction on less vulnerable cases (PURCHASE, TEXAS and LOCATION), while incurring less than $1\%$ test accuracy drop. 


\mypara{Comparison with adv-reg.} We note that the adversarial regularization defense incurs significant test accuracy drop for all cases (detailed numbers are shown in   \Cref{tab:full_def_eval_tpr_cifar10,tab:full_def_eval_tpr_cifar100,tab:full_def_eval_tpr_binary} in the appendix) except when AlexNet is used (for CIFAR-10 or CIFAR-100). However, for these two cases, MIST+Mixup (blue curve) provides much lower PLR than adversarial regularization (purple dot) when similar test accuracy is achieved, as shown in Figure \ref{fig:all_def_comp}. In the adv-reg defense, the model is trained to defend against one attack model that evolves with it in adversarial training, and the resulting model can still be vulnerable to other attacks.  Furthermore, the nature of adversarial training is such that one has to be very careful on tuning the hyperparameters because it is highly likely that the target model will have very low test accuracy. 


\mypara{Comparison with Mem-guard.} The Mem-guard defense is a post-processing defense, which means the model utility (test accuracy in our case) is maintained. From \Cref{fig:all_def_comp}, we can conclude that MIST+Mixup (blue curve) provides lower PLR than Mem-guard (olive dot) when similar test accuracy is achieved. We note that due to computational resources and time constraint, we are not able to evaluate Mem-guard against the CANARY attack, since the CANARY attack involves instance optimization through multiple iterations and for each iteration the Mem-guard defense needs to calculate a new noise regarding each newly perturbed instance. If we only compare the PLR of the LIRA attack, we can see that MIST+Mixup outperforms Mem-guard for all cases. Moreover, the PLR of the CANARY attack against MIST+Mixup is still lower than the PLR of the LIRA attack against Mem-guard for all CIFAR experiments. The Mem-guard defense uses one attack model to generate the noise to be added to each prediction, however this particular noise may not be effective against other attacks. 


\mypara{Comparison with DMP.} In this comparison, we consider the case where no auxiliary unlabeled data is available because in practice collecting extra data is intractable for most privacy-concerning cases, for example medical images. More specifically, we train a GAN using the private training set and draw samples using the trained GAN to train the student model. However in our experiments, we see that the DMP defense also incurs significant test accuracy drop comparing to the no-defence case. We include the detailed numbers in \Cref{tab:full_def_eval_tpr_cifar10,tab:full_def_eval_tpr_cifar100,tab:full_def_eval_tpr_binary} in the appendix.  

\mypara{Comparison with Mixup+MMD.} From \Cref{fig:all_def_comp}, we can conclude that MIST+Mixup (blue curve) provides lower PLR than Mixup+MMD (green dot) when similar test accuracy is achieved. We note that our proposed cross-difference loss is similar to the MMD (mean maximum discrepancy) loss proposed in Li et al.~\cite{li2021membership}. There are two key differences.  One is the granularity, i.e., our proposed cross-difference loss is calculated on instance level, but the MMD loss is calculated on class level, where the loss is defined as the difference between averages of two classes. The other is that in~\cite{li2021membership}, MMD loss is relative to a validation dataset.  Here, we divide the dataset into multiple subsets and the cross-difference loss is relative to results from update from other subsets.  As demonstrated in \ref{sec:ablation}, training using divided subset can help the model generalize better (and less vulnerable to MI attacks) even without using the cross-difference loss.

\mypara{Comparison with SELENA.} We first note that the SELENA defense and its evaluation against shadow-model based attacks require extraordinarily large amount of computational resources. This is because the SELENA defense needs to train $25$ (the default value) models and then train one distilled model from these $25$ models. Moreover, if we want to evaluate the LIRA attack against the SELENA defense and the number of shadow models is $100$, then we need to perform the SELENA training procedure for $100$ times to generate $100$ distilled models and in total we need to train $2600$ models for all our models and datasets, which is unattainable in our experimental scenario where each model is not small. Thus, for our experiments we restrict SELENA defense evaluation to the following scenarios: CIFAR-10 with AlexNet, CIFAR-100 with AlexNet, PURCHASE, TEXAS and LOCATION. From Figure \ref{fig:all_def_comp}, we can conclude that MIST+Mixup defense (blue curve) provides lower PLR and higher test accuracy than SELENA defense (yellow dot). Particularly for the TEXAS dataset, the SELENA defense results in more than $25\%$ test accuracy drop, thus we exclude this data point from Figure \ref{fig:all_def_comp} for clearer comparison. One reason that our defense can outperform SELENA is that, during the training process, there is communication (regularizing each other and averaging parameters) between each submodels, however SELENA does not have communication between submodels. Moreover, this communication takes advantage of the proposed cross-difference loss, which helps to mitigate the overfitting and generalize better, thus is more favorable.

\mypara{Comparison with HAMP.}
For the HAMP defense, we only evaluate the effectiveness of their training-time defense, since the test time defense can be integrated with any training time defenses and the test time defense is approximately equivalent to providing the predicted label only. In \Cref{fig:all_def_comp}, if we compare MIST+Mixup (blue curve) with HAMP (red curve), we can conclude that MIST+Mixup gives better privacy-utility tradeoff than HAMP. For the HAMP defense, our defense considers the MI difficulty of different individual instances and regularize more on vulnerable instances, however the HAMP defense treats all instances in the same fashion by applying an entropy-based regularizer and smoothed label. Moreover, by leveraging subspace learning, we can achieve better model generalization.

\mypara{Comparison with DP-SGD.}
Note that for the DP-SGD defense we tune the hyperparameter so that the model trained with DP-SGD can achieve similar test accuracy and in this case, the privacy budget is huge.  In \Cref{fig:all_def_comp}, if we compare MIST+Mixup (blue curve) with DP-SGD (brown curve), we can conclude that MIST gives better privacy-utility tradeoff than DP-SGD. The DP-SGD defense add random noise drawn from a given gaussian distribution, thus adding (expected) the same amount of noise for each instance, which implicitly treats each instance in the same way. However, our proposed MIST defense is able to treat each instance with different level of protection via the proposed cross-difference loss, thus can outperform the DP-SGD defense.

To summarize, our MIST defense can outperform all existing defenses for all evaluated cases.
In addition, our defense also provides a ``knob'' (the parameters $\lambda$ and $\Ttwo$ of Phase 2) to further fine-tune the model, so practitioners can obtain different tradeoffs between test accuracy and MI attack robustness. This tradeoff flexibility is absent from some of the baseline defenses, unfortunately.


\subsection{Evaluating against label only MIAs}
Now we evaluate the effectiveness of MIST against label-only MIAs from Choquette-Choo et al.~\cite{choquette2021label}. Particularly, we evaluate the more effective adversarial perturbation label-only MIA. The intuition of this label-only MIA is that the training samples need more perturbations to get their predicted class to be flipped. We use both the PLR at FPR$=0.1\%$ and the AUC score to evaluate the effectiveness of the distance attack. The results are shown in \Cref{tab:label_only_eval}. It is easy to see that the label-only MI attack is not effective in detecting most vulnerable instances (PLR close to 1 means that TPR is  close to FPR). Moreover, we can conclude that our proposed defense can significantly reduce the AUC score of the distance attack and outperform all existing attacks in defending against this label only membership inference attack.

\begin{table*}
    \centering
    \begin{tabular}{lrrrr}
        \toprule
        \multirow{3}{3cm}{Defense} &  \multirow{3}{2cm}{Attack PLR (FPR@$0.1\%$) on CIFAR-10} & \multirow{3}{2.2cm}{Attack PLR (FPR@$0.1\%$) on CIFAR-100} & \multirow{3}{2.2cm}{Attack AUC \\ on CIFAR-10} & \multirow{3}{2.2cm}{Attack AUC \\ on CIFAR-100}\\
        & & & & \\
        & & & & \\
        \midrule
          No Defense &   1.10 &   0.90 &   0.641  &  0.826 \\
         Adv-reg &   1.39 &   1.13 &  0.631 &  0.815 \\
         Mem-guard &   1.10 &   0.90 &  0.641 &  0.826  \\
         Mixup+MMD &   1.30 &   1.50 &  0.605 &  0.742\\
         SELENA &   1.21 &   1.11 &  0.542 &  0.545  \\
         HAMP &   1.73 &   1.94 &  0.577 &  0.633\\
         DP-SGD &   1.40 &   1.03 &   0.645 &  0.661\\
        \midrule
         MIST+Mixup (ours) &   \textbf{1.05} &   \textbf{0.70} &  \textbf{0.536} &  \textbf{0.540} \\
        \bottomrule
    \end{tabular}
    \caption{Evaluation against label only MIA. CIFAR dataset, AlexNet. The DMP defense would result in training failure (model not converging), thus excluded from this table.}
    \label{tab:label_only_eval}
\end{table*}

\subsection{Evaluating against white-box MIAs}
Carlini et al.~\cite{carlini2020cryptanalytic} proposed a model extraction method that can accurately extract parameters for a 5-layer fully connected neural network with $1110$ parameters.
If an effective model extraction attack against deep DNNs is discovered, then blackbox access to a target model becomes equivalent to whitebox access, and we need to evaluate defenses against whitebox attacks as well.  
In this subsection, we evaluate all defenses against the white-box MIA from Nasr et al.\cite{nasr2019comprehensive} to understand the effect of our proposed defense against white-box attacks. The results are shown in \Cref{tab:whitebox_eval_cifar10,tab:whitebox_eval_cifar100}.  
One can also see that our proposed defense can provide significantly lower PLR and AUC score comparing to other existing defenses, which shows that our proposed defense can provide benefits against both black-box attacks and white-box attacks.  From the results we can also see that this whitebox attack is no more effective than state-of-the-art blackbox attacks such as LiRA.  This fact may not be as counter-intuitive as it first seems.  Exact model parameters are affected by many factors beyond the training dataset, including initial randomness in initialization, hyper-parameters, and randomness during training (e.g., due to dropout).  A member instance affects the model parameters because the parameters are trained to minimize loss on the instance, thus the loss on the instance (which is used in blackbox attacks) is the main ``footprint'' left the instance being a member.  

\begin{table*}
    \centering
    \begin{tabular}{lrrrrrrrr}
        \toprule
        Defense & No-def & MIST+Mix-up & Adv-reg & Memg-guard & Mix-up+MMD & SELENA & HAMP & DP-SGD  \\
         \midrule
           PLR(FPR@$0.1\%$) &  1.01 &  \textbf{0.00} &  0.00 &  1.01 &  0.00 &  0.97 &  0.40 &  0.33 \\
           AUC score &  0.734 &  \textbf{0.503} &  0.706 &  0.741 &  0.512 &  0.541 &  0.579 &  0.584 \\
        \bottomrule
    \end{tabular}
    \caption{Evaluation against white-box MIA. CIFAR-10 dataset, AlexNet. The DMP defense would result in training failure (model not converging), thus excluded from this table.}
    \label{tab:whitebox_eval_cifar10}
\end{table*}

\begin{table*}
    \centering
    \begin{tabular}{lrrrrrrrr}
        \toprule
        Defense & No-def & MIST+Mix-up & Adv-reg & Memg-guard & Mix-up+MMD & SELENA & HAMP & DP-SGD  \\
        \midrule
          PLR(FPR@$0.1\%$) &  1.61 &  \textbf{0.00} &  1.68 &  1.61 &  0.00 &  2.08 &  3.12 &  0.32 \\
          AUC score &  0.871 &  \textbf{0.505} &  0.866 &  0.871 &  0.542 &  0.541 &  0.663 &  0.667 \\
        \bottomrule
    \end{tabular}
    \caption{Evaluation against white-box MIA. CIFAR-100 dataset, AlexNet. The DMP defense would result in training failure (model not converging), thus excluded from this table.}
    \label{tab:whitebox_eval_cifar100}
\end{table*}

\begin{table*}[]
    \centering
    \begin{tabular}{lrrrrr}
        \toprule
        \multirow{2}{10em}{Dataset-Model} & \multirow{2}{5em}{Defense} &   \multirow{2}{4em}{Acc. on $X_\text{vul}$} &   \multirow{2}{4em}{Test Acc. on $X_\text{test}$} &   \multirow{2}{5em}{PLR (FPR@$1\%$) on $X_\text{vul}$} &   \multirow{2}{5em}{PLR (FPR@$1\%$) on $X_\text{whole}$} \\
        & & & & & \\
        & & & & & \\
        \midrule
          CIFAR-100, AlexNet &  No Defense &  1.00 &  0.30 &  73.89 &  25.65  \\
        
         CIFAR-100, AlexNet &  Adv-reg &  0.77 &  0.32 &  83.31 &  8.18 \\
        
         CIFAR-100, AlexNet &  Mem-guard  &  0.95 &  0.30 &  81.98 &  32.19 \\
        
         CIFAR-100, AlexNet &  Mixup+MMD  &  0.37 &  0.30 &  52.61 &  23.13 \\
        
         CIFAR-100, AlexNet &  HAMP  &  0.30 &  0.31  &  41.45 &  17.00 \\
        
         CIFAR-100, AlexNet &  DP-SGD  &  0.21 &  0.30  &  18.76 &  7.23 \\
        
         CIFAR-100, AlexNet &  { Data Removal} &  0.08 &  0.28 &  N/A &  26.53 \\ 
        \cmidrule(lr){1-6}
        
         CIFAR-100, AlexNet &  MIST+Mixup (ours)  &  0.13 &  0.30 &  \textbf{16.10} &  \textbf{6.76} \\
        \midrule
        \midrule
        
         CIFAR-10, AlexNet &  No Defense  &  1.00 &  0.70 &  31.68 &  18.78\\
        
         CIFAR-10, AlexNet &  Adv-reg  &  0.78 &  0.71  &  27.71 &  43.91 \\
        
         CIFAR-10, AlexNet &  Mem-guard  &  0.95&  0.70 &  31.74 &  6.64 \\
        
         CIFAR-10, AlexNet &  Mixup+MMD  &   0.62&  0.70 &  30.80 &  9.79 \\
        
         CIFAR-10, AlexNet &  HAMP  &  0.51&  0.71 &  16.24 &  5.42\\
        
         CIFAR-10, AlexNet &  DP-SGD  &  0.58&  0.69  &  28.95 &  5.63\\
        
         CIFAR-10, AlexNet &  { Data Removal} &  0.30 &  0.68 &  N/A &  19.11\\ 
        \cmidrule(lr){1-6}
        
         CIFAR-10, AlexNet &  MIST+Mixup (ours)  &  0.41&  0.70 &  \textbf{13.15} &  \textbf{5.26} \\
        \bottomrule
    \end{tabular}
    \caption{Evaluation against label only MIA. CIFAR dataset, AlexNet. The DMP defense would result in training failure (model not converging), thus excluded from this table. The SELENA defense is also removed for a fair comparison. $X_\text{vul}$ stands for the vulnerable instance set (of size 1000). $X_\text{test}$ stands for the whole test set. $X_\text{whole}$ stands for the whole evaluation set that we use in previous experiments, which includes $20,000$ training instances. For No Defense case, the $X_\text{vul}$ is included in training, thus the test accuracy on $X_\text{vul}$ is not applicable. For the PLR metric on $X_\text{vul}$, since this is not part of the training set for the data removal defense, this metric is not applicable.}
    \label{tab:outlier_eval}
\end{table*}

\subsection{Comparing against other baselines on the most vulnerable instances}
In this subsection, we want to understand the impact of all defenses over the most vulnerable instances in the training set. For this experiment we consider AlexNet, CIFAR-100 and CIFAR-10 datasets. Our first step is finding the set of most vulnerable instances, which we will denote $X_\text{vul}$. The size of $X_\text{vul}$ is set to be $1000$ in this experiment. We train $100$ models using AlexNet on CIFAR-100 (and CIFAR-10 datasets) without any defense and perform LIRA attack. Recall that the LIRA attack would produce two normal distributions of the loss (one for member case and one for non-member case) for each instance. For each instance, we calculate the non-overlapping area of these two normal distributions and choose the top $1000$ instances with the largest non-overlapping area to construct the $X_\text{vul}$. Intuitively, the larger the non-overlapping area is, the more vulnerable the instance is against membership inference attacks. 

Using the $X_\text{vul}$ above we can evaluate two metrics: the test accuracy on $X_\text{vul}$ and the highest attack PLR on $X_\text{vul}$. Moreover, we add the highest attack PLR on the whole training set for comparison and further illustration in \Cref{tab:outlier_eval}. Note that for a fair comparison between different defenses (defenses evaluated in this subsection should result in less than $1\%$ test accuracy drop) and clearer results presentation, we remove the SELENA defense and the DMP defense from this experiment, since these two defenses would result in significant test accuracy drop. In conclusion, the results in \Cref{tab:outlier_eval} show that our MIST defense can outperform all existing membership inference defenses in terms of attack effectiveness reduction on both the vulnerable instances $X_\text{vul}$ and on the whole MI evaluation dataset $X_\text{whole}$.

In particular, a natural defense is the removal from training of the most vulnerable instances against membership inference attacks. The intuition is that, once the most vulnerable instances are removed, the remaining training instances should be more robust against MI attacks. 
Unfortunately, our experiments show that this defense is surprisingly ineffective, with re-identification metrics close to the ``No Defense'' scenario, which is also detailed in Table \ref{tab:outlier_eval}.

The difference between MIST and the data removal defense is worth a closer look. For the test accuracy on $X_\text{vul}$, the performance of the data removal defense is significantly worse than MIST defense. For the PLR at $1\%$ FPR on $X_\text{vul}$, since the whole $X_\text{vul}$ is removed for the data removal defense, this PLR metric is not applicable to the data removal defense and there is no privacy threat against $X_\text{vul}$. 
However, for the PLR at $1\%$ FPR on $X_\text{whole}$, the data removal defense performs much worse than MIST defense and close to the ``No Defense'' scenario, which makes the data removal defense similar to having no defense. 

Let's take CIFAR-100, AlexNet as one example. Intuitively, when FPR=$1\%$, the PLR is $25.65$ for no-defense case, which means that $0.01\times25.65\times20000=5130$ instances have been detected as members. If we remove the $X_\text{vul}$ from training, then the expected PLR at FPR=$1\%$ should be $\frac{4130}{5130}*25.65=20.65$, however, the resulted PLR is now $26.53$. This result implies that even though the vulnerable instances $X_\text{vul}$ are removed from the training set, now other instances became vulnerable against MI attacks.
One intuitive explanation is that, since $X_\text{vul}$ is mostly constructed by outliers (since outliers are harder to fit and thus easier to be attacked by membership inference attacks), removing them from training would make some other instances to become outliers, just like peeling an onion. This finding is aligned with the recent findings of Carnili et al.~\cite{carlini2022privacy}, which show that removing vulnerable instances from the training set allows other instances to become outliers and thus
vulnerable. 

To summarize, the data removal defense seems to be ineffective. Moreover, we show that MIST, our proposed defense, can provide the most reduction in attack effectiveness on the most vulnerable set, while achieving the best average case test accuracy and (both average case and extreme case) privacy tradeoff of all baselines.

\subsection{Choosing hyperparameters} Now we discuss how to determine the two most crucial hyperparameters of our proposed defense: the number of models $C$ and the weight of the cross-difference loss. In this paper, we use the following method to determine these two hyperparameters: we choose the number of models $C$ to achieve the highest validation accuracy possible and then choose the weight of the cross-difference loss to provide the strongest defense while making sure the test accuracy drop is less than $1\%$ comparing to the base case without any defense. In Figure \ref{fig:num_user_nomixup} in the Appendix (with more detailed results in \Cref{tab:num_user,tab:num_user_cifar100} in the Appendix), we show that the customized subspace learning (without the cross-difference loss) can help the model generalize better, thus we would like to use the model with the highest validation accuracy. Moreover, the mixup data augmentation can also help to further improve the validation accuracy. Hence, subspace learning and mixup combined get higher test accuracy boost compared to standard SGD, which then gives us room to increase the regularization of the cross difference loss without lower test accuracy when compared against standard SGD. 

\begin{figure}
    \centering
    \includegraphics[height=4.5cm,width=0.5\textwidth]{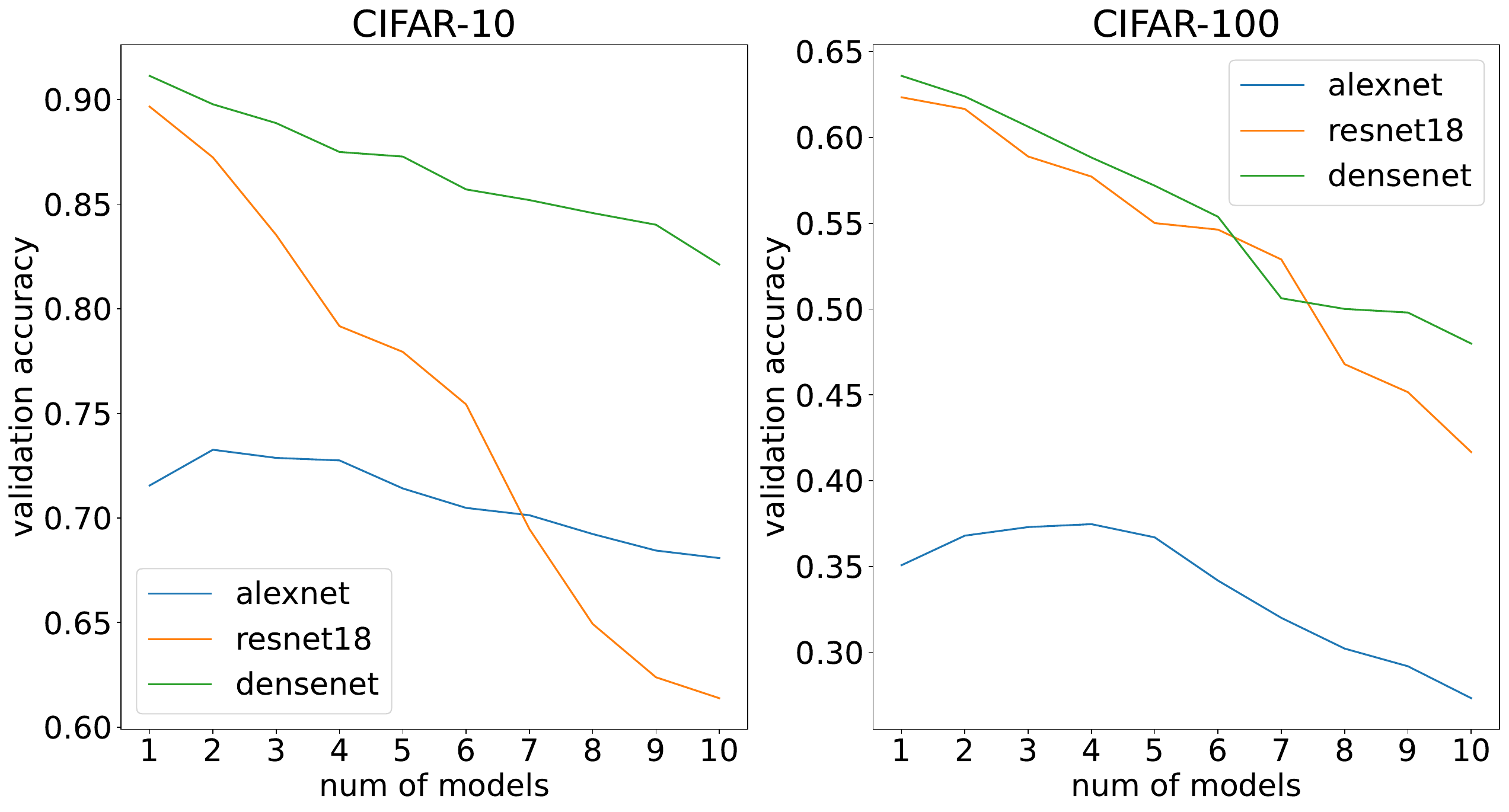}
    \vspace{-0.1cm}
    \caption{Validation accuracy v.s. number of models $C$. Mixup data augmentation is applied.}
    \vspace{-0.4cm}
    \label{fig:opt_num_user}
\end{figure}

In \Cref{fig:opt_num_user}, we show the validation accuracy curve v.s. the number of models $C$ for all models and datasets where mixup data augmentation is applicable. Recall that we need to choose the number of models $C$ that achieves the highest validation accuracy. However, we also need to have at least two models, so that our cross-difference loss is applicable. Thus, where mixup data augmentation is applicable, for all models and datasets except CIFAR-100 and AlexNet, the number of models $C$ should be $2$. For the CIFAR-100 and AlexNet, the number of models $C$ should be four. For the case where the mixup data augmentation is not applicable (TEXAS dataset and PURCHASE dataset), we choose to use 4 models for TEXAS dataset and 4 models for PURCHASE dataset based on \Cref{fig:num_user_nomixup} in the Appendix. In practice, we recommend developers to generate this validation accuracy curve for their own datasets to determine the number of models $C$. If the computational resources are limited, our ad-hoc advice is to use $2$ to $4$ models since this range was suitable for all our datasets.

As for the choice of the hyperparameter $\lambda$ of Phase 2 (the cross-difference loss minimization), we choose $\lambda$ as to provide the strongest defense while making sure the test accuracy drop is less than $1\%$. In practice, the training does not need to follow this procedure, as $\lambda$ may be utilized as a knob to tune a privacy-utility tradeoff, which allow practitioners to determine this tradeoff based on their needs.

\subsection{Ablation Study}
\label{sec:ablation}

\begin{figure}[t]
    \centering
    \includegraphics[height=10cm,width=0.5\textwidth]{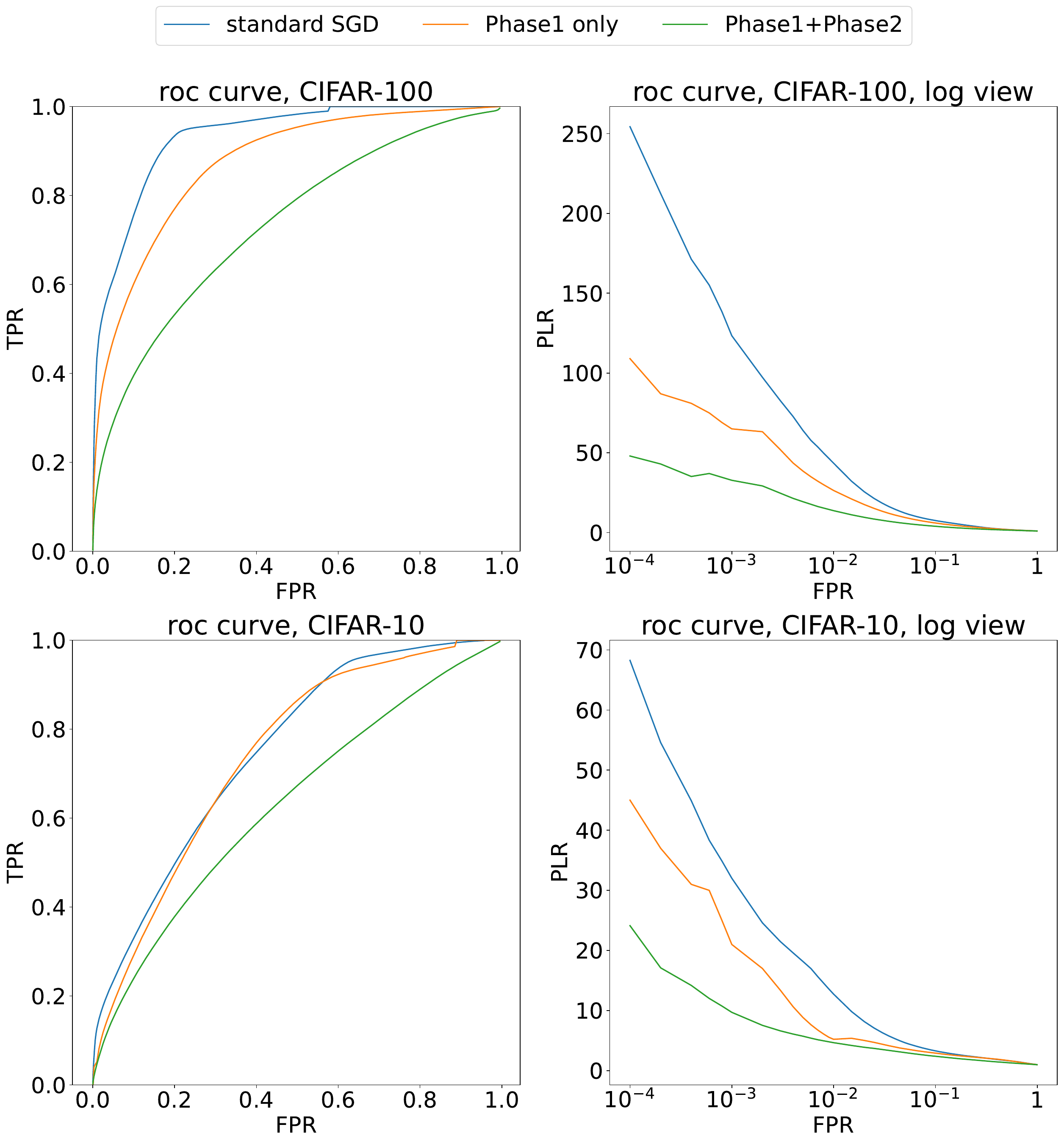}
    \caption{The clear positive impact of Phase 1 and Phase 2. Impact measured on ROC curve and PLR at low FPR region. LIRA attack is evaluated to produce the ROC curve and PLR curve. CIFAR-100 and CIFAR-10 using AlexNet. }
    \vspace{-0.4cm}
    \label{fig:cross-diff-impact}
\end{figure}

\mypara{The positive impact of subspace learning and cross-difference loss.} In this ablation study, we use AlexNet on CIFAR-10 and CIFAR-100 dataset. For CIFAR-10 dataset, we use $3$ models and for CIFAR-100 dataset we use $7$ models. These two values are chosen because the highest test accuracy (higher than the standard SGD case with one model) can be achieve with this choice when no cross-difference loss and mix-up data augmentation is applied (detailed results are shown in \Cref{tab:num_user,tab:num_user_cifar100} in the Appendix). For the $\lambda$ of the cross difference loss, we choose the $\lambda$ so that the maximum defense can be provided while the test accuracy drop is less than $1\%$ in this experiment. We compare the attack effectiveness against model trained without defense, trained with Phase 1 only and trained with both Phases 1 and 2. The results are shown in \Cref{fig:cross-diff-impact}. We can conclude that the only performing Phase 1 of our MIST algorithm (ignoring Phase 2) can still be used as a defense which can preserve utility and decrease attack effectiveness at the same time with appropriate number of models $C$. We also repeat this experiment on other models and datasets and we observe similar phenomenon. Moreover, adding Phase 2 (cross-difference loss minimization) can provide further robustness against the LIRA attack.

\mypara{Different cross-difference loss implementations.}
In this ablation study, we experiment on different implementations for the cross-difference loss. More specifically, we use three different implementations: KL-divergence, \Cref{eq:cross_diff} and $L_2$ version of \Cref{eq:cross_diff}. We run experiments using AlexNet on CIFAR-100 dataset and we choose the optimal weights that we can find for each loss function so that the test accuracy drop is less than $1\%$ comparing to the case without any defense. The results are shown in \Cref{tab:loss_func_comp}, where we can conclude that the $L_1$ variant slightly outperforms the other two variants and the KL-divergence variant performs the worst. One possible reason for the KL-divergence variant being worse than $L_1$ is because the KL-divergence considers the distance between two prediction vectors, while the state-of-the-art attacks only focus on the probability of the correct label. Moreover, we also observe that this KL-divergence variant makes the model training unstable, meaning that sometimes the model fails to converge, thus we recommend the use of $L_1$.

\begin{table}[]
    \centering
    \resizebox{.97\columnwidth}{!}{
    \begin{tabular}{l rr}
    \toprule
    Loss function & Test accuracy & PLR at $0.1\%$ FPR\\
    \midrule
    L1    &  0.3049 & 17.43 \\
    L2    & 0.2997 & 17.54 \\
    KL-divergence & 0.3003 & 19.72 \\
    \bottomrule
    \end{tabular}
    }
    \vspace{0.3cm}
    \caption{Comparing different implementations of the MIST loss. CIFAR-100 dataset on AlexNet.}
    \label{tab:loss_func_comp}
\end{table}

\section{Conclusion}
\label{sec:conclusion}

In this work we introduce Membership-Invariant Subspace Training (\EiSL), a method for training classifiers that acts as a defense designed to specifically defend against blackbox membership inference attacks on the most vulnerable instances in the training data (which also helps defending against overall blackbox MI attacks). \EiSL uses a novel two-phase subspace learning procedure that trains models regularized to be {\em membership invariant} (counterfactually invariant to the membership of each instance in the training data $D$). 
Through our proposed two-phase subspace learning method, and a new proposed loss ({\em cross-difference loss}), we can efficiently regularize neural network training towards membership-invariant classifiers. Our experiments empirically show that our trained classifiers are significantly less vulnerable against blackbox membership inference attacks than baselines, especially on those most vulnerable training examples. 

\section{Limitation and Future Works}
\label{sec:future}
The main limitation of the proposed MIST defense is the computational overhead of maintaining multiple models. This is because at each epoch, the MIST defense requires maintaining multiple copies of the central model, performing local model updates, update local models based on predictions from other models, and averaging the model weights. More specifically, if the training set is divided into $C$ subsets, this results in a memory overhead by a factor $C$. One possible mitigation strategy is to perform the training for each submodel in a sequential way and only load one model into memory. However, this solution would result in a time overhead of a factor $C$. Another interesting future direction is adapting the MIST to regression and other non-classification tasks. Finally, there may be other, yet unknown, ways to apply the cross-difference loss that may reduce overhead and improve performance (better privacy-utility tradeoff).

\section{Acknowledgements}

This work was funded in part by the National Science Foundation (NSF) Awards CAREER IIS-1943364, CCF1918483, CNS-2247794, and IIS-2229876, Amazon Research Award, AnalytiXIN, the Wabash Heartland Innovation Network (WHIN), Ford, and CISCO. Any opinions, findings and conclusions or recommendations expressed in this material are those of the authors and do not necessarily reflect the views of the sponsors.
\bibliographystyle{plain}
\bibliography{privacy,ml}
\newpage
\appendix

\section{Further Related Work}
\label{sec:related}



\mypara{Other MI attacks that are similar to LIRA but weaker.} 
Several MI attacks are similar to LIRA and are shown in ~\cite{carlini2022membership} to be less effective in LIRA; we thus do not consider them in the experiments.  We list these attacks below.

Long et al.~\cite{long2017towards} proposed to train instance-specific MI classifiers.
For each instance $x$, there will be two sets of models: trained with $x$ and trained without $x$. Then, the average prediction of $x$ from the member set and the average prediction of $x$ from the non-member set is calculated. $x$ is predicted to be a member if the target model's prediction of $x$ is more similar to the average prediction of the member set (in the sense of KL divergence).


Ye et al.~\cite{ye2022enhanced} followed the same shadow model procedure as ~\cite{watson2021importance} to produce a collection of loss for one instance when this instance is not used in the training. Next, a one-sided hypothesis testing is proposed to predict membership. The advantage of this hypothesis testing is that the attacker could select a false positive rate. However, the possible false positive rate range is limited by the number of shadow models. 

Watson et al.~\cite{watson2021importance} propose to use the shadow models to set a per-instance loss threshold by the average of loss of the instance on shadow models that are not trained using this example. 


\mypara{Other membership inference defenses.} In ~\cite{shokri2017membership}, it was proposed to reduce the information given by the prediction vectors, such as providing only the top-k probabilities and using high temperature in softmax.  This has limited effectiveness as the top-k probabilities give enough information needed by the best attacks, and high temperatures change only the absolute values of confidences, but not the fact that confidences for members tend to be higher than that for non-members. Moreover, this defense is not effective against LIRA, thus we do not consider this defense in our experiments.



\mypara{Membership inference attacks in other settings.} Melis et al.~\cite{melis2019exploiting} identified membership leakage when using FL for training a word embedding function, which is a deterministic function that maps each word to a high-dimensional vector. Given a training batch (composed of sentences), the gradients are all $0$'s for the words that do not appear in this batch, and thus the set of words used in the training sentences can be inferred.  The attack assumes that the participants update the central server after each mini-batch, as opposed to updating after each training epoch.

Nasr et al.~\cite{nasr2019comprehensive} use gradient updates as feature vectors and train an auto-encoder to generate a single-number embedding to predict membership. This attack was also applied in the white-box setting, for which it performs worse than attacks that directly use gradient norm to predict membership, which show only small improvement over blackbox attacks. Another interesting idea in~\cite{nasr2019comprehensive} is that a malicious server can actively improve MI attacks, by applying a gradient ascent with respect to the target instance.  If the instance is a member, then the gradients tend to be larger in order to compensate for the malicious gradient ascent.  

Li et al.~\cite{li2023effective} proposed a new membership inference attack against the federated learning setting. They observed that the cosine similarity between gradients of each data instance and the parameter updates sent by each client has very different distribution for members and non-members. Thus, the cosine similarity is used as the metric to predict members. They also proposed to use the gradient difference between the gradients of each data instance and the parameter updates to predict membership. Their results show that the new proposed attacks can achieve higher TPR at low FPR than the attack proposed in ~\cite{nasr2019comprehensive}. 

Song et al.~\cite{song2019membership} evaluated how adversarial training would affect the privacy leakage. Experiments showed that adversarial training would boost the performance of instance loss membership inference attack and the testing accuracy of the target model will be decreased. 

Liu et al.~\cite{liu2021encodermi} evaluated the privacy leakage of pre-trained models which are trained using unsupervised contrastive learning strategy. The authors proposed a specific membership inference attack against pre-trained models. Given one instance, this new attack generates many perturbed versions of this instance and gather all the embeddings of these perturbed versions using the pre-trained model. The intuition is if one instance is used in the training of this pre-trained model, then the embeddings of its perturbed versions are generally closer to each other. He et al.~\cite{he2021quantifying} utilized the same contrastive learning idea and proposed to fine-tune the pre-trained model to get the final target model. Experiments showed that using pre-trained model as feature extractor would reduce privacy leakage, comparing to training models from scratch. 

Hidano et al.~\cite{hidano2021transmia} evaluated MI attack under the setting where the attacker can know the parameters of some shallow layers. In the experiment, the author assumed that the attacker can get all but the last layer and the parameters of these layers are used to initialize shadow models to facilitate the class-vector attack. With these known parameters, the class-vector attack can outperform its black-box version.

\mypara{Other privacy-threatening attacks.}
Salem et al.~\cite{salem2020updates} studied the possible information leakage of an update set when the machine learning model is updated using the update set. To detect the information leakage, the authors proposed two different attacks: label inference attack and instance reconstruction attack. These two attacks can be applied to both single instance update set case and multi-instances update set case, with only black-box access to the machine learning model.


Zhu et al.~\cite{zhu2020deep} presented a gradient-based instance reconstruction attack. If the gradients of one specific instance are revealed to one adversary who can access the trained model, then the adversary is able to reconstruct the specific instance with high fidelity. One random instance is gradually optimized by matching its gradients with the provided gradients of the specific instance. 




\section{Additional Ablation Study}

\begin{figure}[!t]
    \centering
    \includegraphics[height=16cm,width=0.48\textwidth]{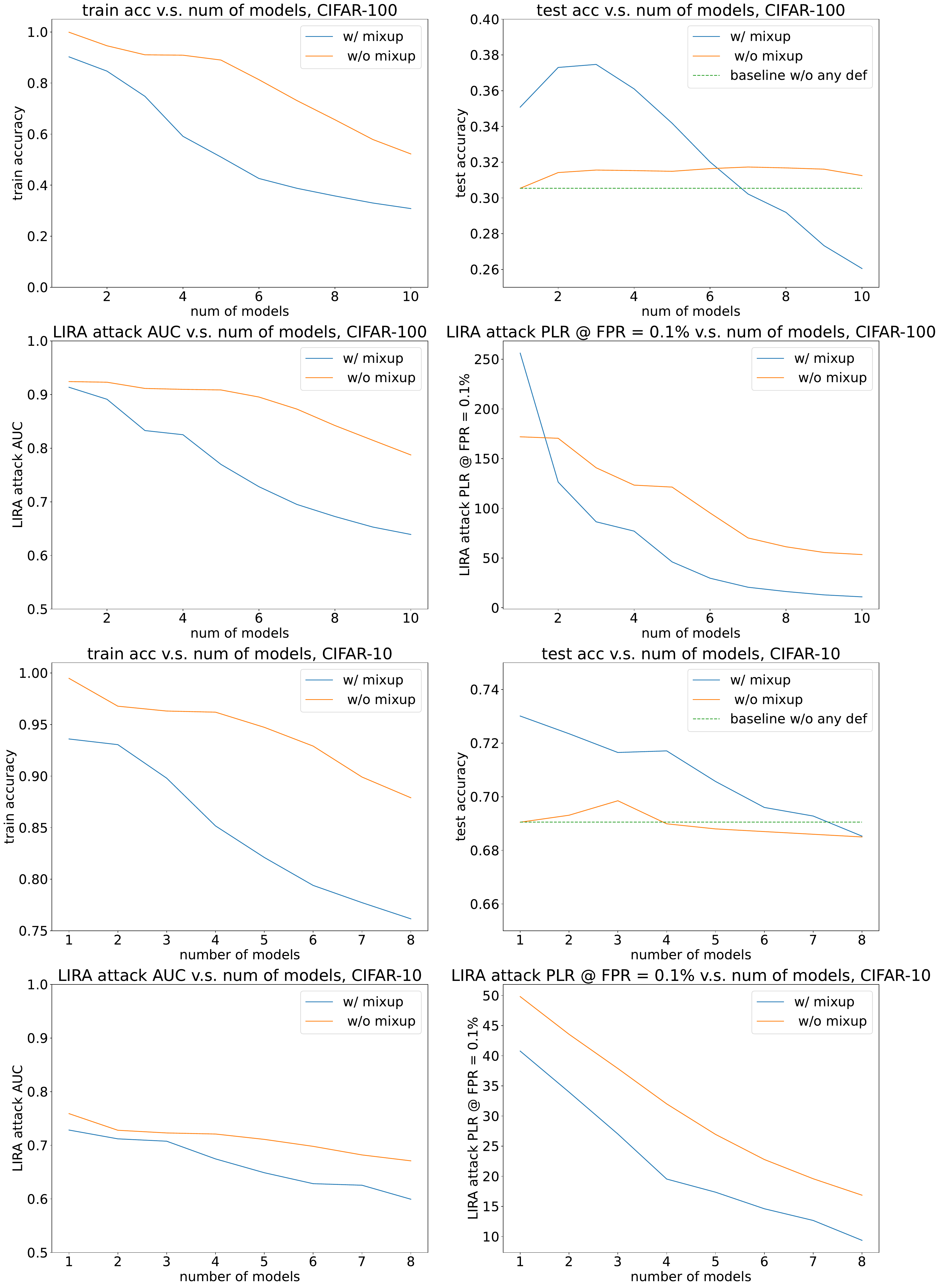}
    \caption{The impact of mixup data augmentation on training accuracy, testing accuracy and attack effectiveness when the number of models is varied. AlexNet, CIFAR-10 and CIFAR-100 datasets.}
    \vspace{-0.4cm}
    \label{fig:num_user_mixup}
\end{figure}

~\\


\mypara{The positive impact of mixup data augmentation.} The experiment is performed using CIFAR-10 and CIFAR-100 dataset on AlexNet. We perform experiments while varying the number of models $K$ from 1 to 8 for CIFAR-10 (10 for CIFAR-100). $D$, which contains 20,000 training instances, is divided into $K$ equal-size, disjoint partitions, and each model will be trained using one partition for one epoch, before we average the model. For each number of models choice, we experiment with two cases: with mixup and without mixup.  The results are shown in \Cref{fig:num_user_mixup}. The detailed numbers are included in Table \ref{tab:num_user}, \ref{tab:num_user_cifar100}, \ref{tab:num_user_mixup} and \ref{tab:num_user_mixup_cifar100} in the appendix. From \Cref{fig:num_user_mixup}, for CIFAR-100 dataset, the first thing we can observe is that the training accuracy is decreased and the testing accuracy is increased when mixup is applied for the same number of models. Furthermore, by adding the mixup data augmentation, the LIRA attack AUC and LIRA attack PLR at low FPR are both reduced when the same number of models are created. This validates the effectiveness of the mixup data augmentation. However, we also noticed that the testing accuracy is always decreasing while creating more than one models and when the number of models is seven, the testing accuracy is very similar to the testing accuracy when there is no defense. Thus, there is a privacy-utility tradeoff that should be considered by the model trainer and the model trainer should choose the number of models based on their own needs. In addition, in order to regularize the trained model with the cross-difference loss, there should be at least two models. For CIFAR-10 dataset, we also notice that the training accuracy is decreased and the testing accuracy is increased when mixup is applied for the same number of models, which again verifies the effectiveness of the mixup data augmentation. The only difference is that, the testing accuracy increases and then decreases while having more models. Therefore, we suggest the model trainers to generate similar figures like Figure \ref{fig:num_user_mixup} and choose the number of models based on their needs. 

\section{Additional Experimental Details and Results}\label{sec:appendix}

\mypara{Hyperparameters.}  In \Cref{tab:training_recipe}, we present all the hyperparameters used in our evaluation. 

\begin{figure*}
    \centering
    \includegraphics[height=19cm,width=0.9\textwidth]{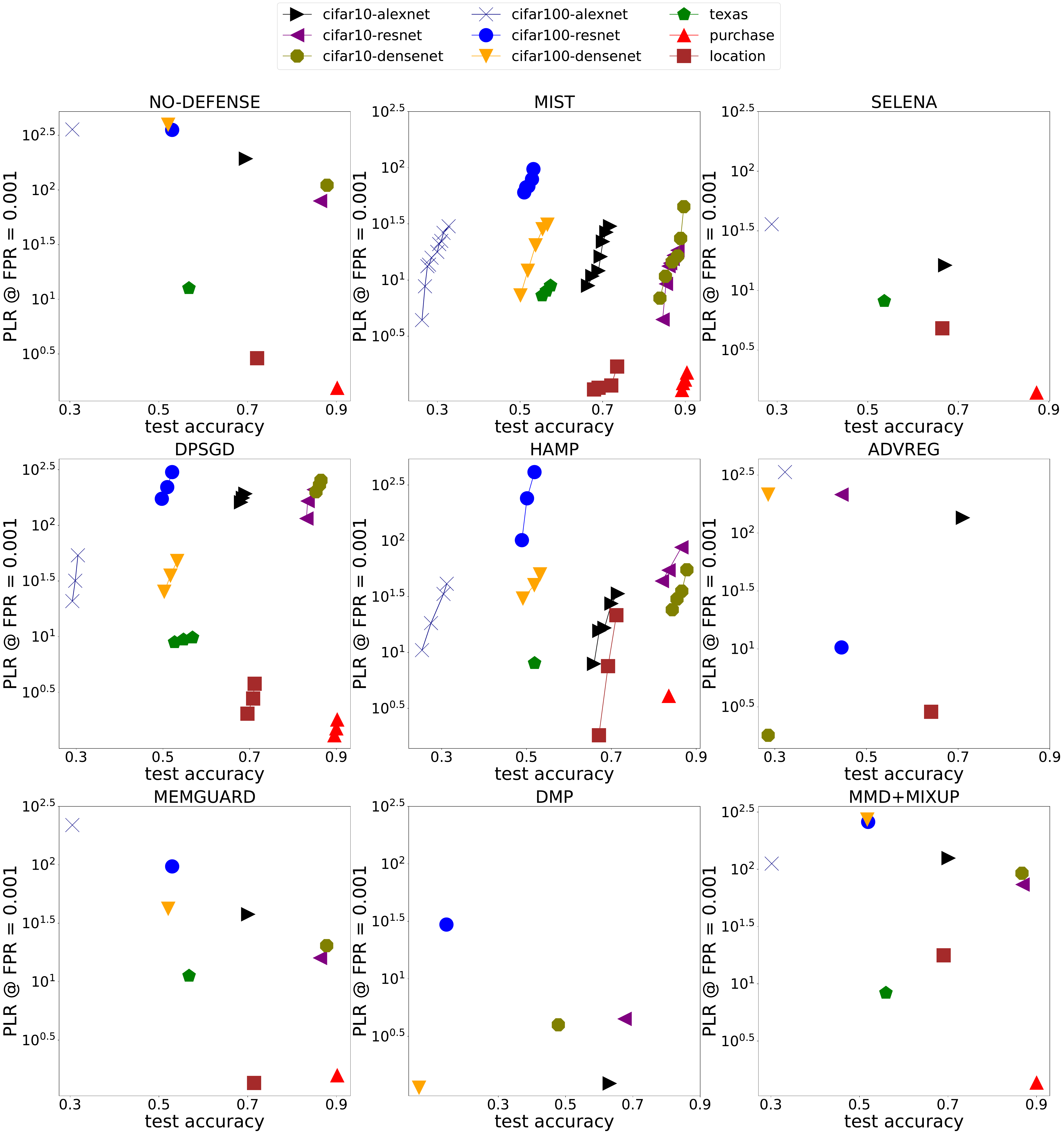}
    \caption{Comparing all defenses using the highest MI attack PLR @ 0.001 FPR among all evaluated attacks. We show the results for each defense in one sub-figure. Notice that for PURCHASE, TEXAS and LOCATION datasets the mixup data augmentation is not applied. }
    \label{fig:all_def_comp_de_wise}
\end{figure*}

\mypara{Detailed numbers for all defenses evaluation.}   In Table \ref{tab:full_def_eval_tpr_cifar10}, \ref{tab:full_def_eval_tpr_cifar100} and \ref{tab:full_def_eval_tpr_binary} we present the detailed PLR numbers at $0.1\%$ FPR for the evaluation of all defenses against all evaluated blackbox attacks. In Table \ref{tab:full_def_eval_auc_cifar10}, \ref{tab:full_def_eval_auc_cifar100} and \ref{tab:full_def_eval_auc_binary} we present the detailed AUC numbers for the evaluation of all defenses against all evaluated blackbox attacks. Moreover, we present the detailed PLR numbers at $1\%$ and $0.5\%$ FPR (in table \ref{tab:full_def_eval_tpr_cifar10_fpr001}, \ref{tab:full_def_eval_tpr_cifar100_fpr001}, \ref{tab:full_def_eval_tpr_binary_fpr001},  \ref{tab:full_def_eval_tpr_cifar10_fpr0005}, \ref{tab:full_def_eval_tpr_cifar100_fpr0005} and \ref{tab:full_def_eval_tpr_binary_fpr0005}) to give the reader a more comprehensive understanding of the defense performance. The main message from the tables above is to show that our proposed MIST defense can provide better privacy-utility tradeoff than existing defenses. In other words, at the same level of testing accuracy, our proposed defense can achieve the best defense result. We also present the Figure \ref{fig:all_def_comp} in a defense-wise way as shown in Figure \ref{fig:all_def_comp_de_wise} for further comparison.

\begin{table*}[tp]
        \normalsize
        \centering
        \begin{tabular}{llrrrrrr}
            \hline
            \vspace{0.05cm}
             dataset & model & lr & epochs & schedule & batch size & number of models & lambda \\
            \hline
            \hline
             CIFAR-10 & AlexNet & 0.05 & 120 & [100] & 100 & 2 & 3.5\\
             CIFAR-10 & ResNet-18 & 0.10 & 300 & [200,250] & 100 & 2 & 3.5\\
             CIFAR-10 & DenseNet\_BC(100,12) & 0.10 & 500 & [300,350] & 100 & 2 & 6.0\\ 
             CIFAR-100 & AlexNet & 0.05 & 120 & [100] & 100 & 4 & 12.0 \\
             CIFAR-100 & ResNet-18 & 0.10 & 300 & [200,250]  & 100 & 2 & 13.0 \\
             CIFAR-100 & DenseNet\_BC(100,12) & 0.10 & 500 & [300,350] & 100  & 2 & 13.0 \\ 
             PURCHASE-100 & PURCHASE-100 & 0.10 & 100 & None & 100 & 4 & 40.0 \\
             TEXAS-100 & TEXAS-100 & 0.10 & 100 & None & 100 & 4 & 25.0 \\
             LOCATION &  LOCATION & 0.10 & 100 & None & 100 & 4&  14.0 \\
            \hline
        \end{tabular}
        \vspace{0.3cm}
        \caption{Training recipe for different models. Learning rate is adjusted to 0.1x when current epoch is in schedule.}
        \label{tab:training_recipe}
\end{table*}

\begin{table*}[!t]
    \centering
    \begin{tabular}{lrrrrrrrrrr}
        \toprule
        \multirow{8}{3em}{\centering Dataset-Model} & \multirow{8}{3em}{\centering Defense} & \multirow{8}{3em}{\centering Train Acc.} & \multirow{8}{3em}{\centering Test Acc.}  & \multirow{8}{4em}{ \centering random-perturbation Attack PLR$\downarrow$\\ (FPR @ $0.1\%$) ~\cite{jayaraman2021revisiting}}  & \multirow{8}{3em}{ \centering LOSS Attack PLR$\downarrow$\\ (FPR @ $0.1\%$) ~\cite{yeom2018privacy}} & \multirow{8}{3em}{ \centering modified-entropy Attack PLR$\downarrow$\\ (FPR @ $0.1\%$) ~\cite{song2019membership}}  & \multirow{8}{3em}{ \centering class-NN Attack PLR$\downarrow$\\ (FPR @ $0.1\%$) ~\cite{shokri2017membership}} & \multirow{8}{3em}{\centering LIRA Attack \\ PLR$\downarrow$\\ (FPR @ $0.1\%$)~\cite{carlini2022membership}}  & \multirow{8}{3em}{\centering CANARY Attack PLR$\downarrow$ \\ (FPR @ $0.1\%$)~\cite{wen2022canary}} \\
         & & & & & & \\
          & & & & & & \\
          & & & & & & \\
          & & & & & & \\
          & & & & & & \\
          & & & & & & \\
          & & & & & & \\
          & & & & & & \\
         \midrule
         \midrule
        CIFAR-10, AlexNet & No-def & 0.9989 & 0.6960 & 0.00 & 0.00 & 0.00 & 33.19 & 49.81 & \textbf{193.10} \\
        CIFAR-10, AlexNet & Adv-reg & 0.9781 & 0.7103 & 0.00 & 0.00 & 0.00 & 21.88 & 12.18 & \textbf{135.38} \\
        CIFAR-10, AlexNet & Mem-guard & 0.9930 & 0.7010 & 0.00 & 0.15 & 0.15 & 4.67 & \textbf{37.67} & OOT \\
        CIFAR-10, AlexNet & DMP & 0.4141 & 0.4043 & 0.00 & 0.38 & 0.50 & 1.07 & 1.23 & 0.99 \\
        CIFAR-10, AlexNet & Mixup+MMD & 0.8805 & 0.7012 & 0.00 & 0.10 & 0.12 & 1.25 & 26.29 & \textbf{125.18}\\
        CIFAR-10, AlexNet & SELENA & 0.7250 & 0.6710 & 0.00 & 0.22 & 0.30 & 1.49 & 6.38 & \textbf{16.25} \\
        CIFAR-10, AlexNet & HAMP & 0.8615 & 0.7106 & 0.00 & 0.12 & 0.12 & 1.06 & 10.15 & \textbf{33.58} \\
        CIFAR-10, AlexNet & DP-SGD & 0.9881 & 0.6925 & 0.00 & 0.10 & 0.10 & 192.10 & 29.03 & \textbf{103.50}\\
        CIFAR-10, AlexNet & MIST+Mixup & 0.7589  & 0.7006 & 0.00 & 0.10 & 0.10 & 3.36 & 11.82 & \textbf{21.93}\\
        \midrule
        CIFAR-10, ResNet18 & No-def & 0.9991 & 0.8631 & 0.00 & 0.00 & 0.00 & 15.30 & 23.99 & \textbf{79.30} \\
        CIFAR-10, ResNet18 & Adv-reg & MNC & MNC & MNC & MNC & MNC & MNC & MNC & MNC\\
        CIFAR-10, ResNet18 & Mem-guard & 0.9999 & 0.8630 & 0.00 & 0.00 & 0.00 & 4.61 & \textbf{15.96} & OOT \\
        CIFAR-10, ResNet18 & DMP & 0.6973 & 0.6745 & 0.00 & 0.49 & 0.59 & 1.19 & 1.37 & 4.48 \\
        CIFAR-10, ResNet18 & Mixup+MMD & 0.9630 & 0.8685 & 0.00 & 0.05 & 0.10 & 4.58 & 12.07 & \textbf{73.60} \\
        CIFAR-10, ResNet18 & SELENA & OOT & OOT & OOT & OOT & OOT & OOT & OOT & OOT  \\
        CIFAR-10, ResNet18 & HAMP & 0.9957 & 0.8651 & 0.00 & 0.00 & 0.00 & 2.65 & 22.61 & \textbf{87.38}\\
        CIFAR-10, ResNet18 & DP-SGD & 0.9797 & 0.8719 & 0.00 & 0.00 & 0.00 & 254.81 & 6.02 & \textbf{34.78} \\
        CIFAR-10, ResNet18 & MIST+Mixup & 0.8802 & 0.8633 & 0.00 & 0.10 & 0.10 & 1.39 & 4.81 & \textbf{14.13 }\\
        \midrule
        CIFAR-10, DenseNet & No-def & 0.9995 & 0.8788  & 0.00 & 0.00 & 0.00 & 291.60 & 30.11 & \textbf{110.49} \\
        CIFAR-10, DenseNet & Adv-reg & 0.2891 & 0.2841 & 0.00 & 0.00 & 0.00 & 0.54 & 0.89 & 1.01 \\
        CIFAR-10, DenseNet & Mem-guard & 0.9999 & 0.8779 & 0.00 & 0.00 & 0.00 & 9.52 & \textbf{20.34} & OOT \\
        CIFAR-10, DenseNet & DMP & 0.4845 & 0.4785 & 0.00 & 0.38 & 1.17 & 2.72 & 3.49 & 3.89 \\
        CIFAR-10, DenseNet & Mixup+MMD & 0.9577 & 0.8671 & 0.00 & 0.10 & 0.10 & 48.31 & 27.31 & \textbf{92.34}\\
        CIFAR-10, DenseNet & SELENA & OOT & OOT & OOT & OOT & OOT & OOT & OOT & OOT  \\
        CIFAR-10, DenseNet & HAMP & 0.9639 & 0.8780 & 0.00 & 0.10 & 0.19 & 6.34 & 19.15 & \textbf{55.12}\\
        CIFAR-10, DenseNet & DP-SGD & 0.9898 & 0.8644 & 0.00 & 0.00 & 0.00 & 253.19 & 7.99 & \textbf{27.69} \\
        CIFAR-10, DenseNet & MIST+Mixup & 0.9097 & 0.8814 & 0.00 & 0.10 & 0.15 & 1.62 & 5.21 & \textbf{16.38} \\
        \bottomrule
    \end{tabular}
    \caption{Defense evaluation on CIFAR-10. OOT stands for out-of-time. The SELENA defense cannot be evaluated on ResNet18 and DenseNet because this defense requires more than half an year to run the experiments. The Mem-guard defense cannot be evaluated against CANARY attack also because of time constraint. MNC stands for Model Not Converging (training failure). DMP defense and Adv-reg defense will sometimes result in extremely low testing accuracy and thus the attack evaluation against these defenses is not meaningful. Lower PLR while preserving test accuracy is better.}
    \label{tab:full_def_eval_tpr_cifar10}
\end{table*}

\begin{table*}[!t]
    \centering
    \resizebox{1.02\textwidth}{!}{
    \begin{tabular}{lrrrrrrrrrr}
        \toprule
        \multirow{8}{7em}{\centering Dataset-Model} & \multirow{8}{3em}{\centering Defense} & \multirow{8}{3em}{\centering Train Acc.} & \multirow{8}{3em}{\centering Test Acc.}  & \multirow{8}{4em}{ \centering random-perturbation Attack PLR$\downarrow$\\ (FPR@ $0.1\%$) ~\cite{jayaraman2021revisiting}}  & \multirow{8}{3em}{ \centering LOSS Attack PLR$\downarrow$\\ (FPR@ $0.1\%$) ~\cite{yeom2018privacy}} & \multirow{8}{3em}{ \centering  modified-entropy Attack PLR$\downarrow$\\ (FPR@ $0.1\%$) ~\cite{song2019membership}}  & \multirow{8}{3em}{ \centering class-NN Attack PLR$\downarrow$\\ (FPR@ $0.1\%$) ~\cite{shokri2017membership}} & \multirow{8}{3em}{\centering LIRA Attack \\ PLR$\downarrow$\\ (FPR@ $0.1\%$)\cite{carlini2022membership}}  & \multirow{8}{3em}{\centering CANARY Attack PLR$\downarrow$ \\ (FPR@ $0.1\%$)\cite{wen2022canary}} \\
          & & & & & & \\
          & & & & & & \\
          & & & & & & \\
          & & & & & & \\
          & & & & & & \\
          & & & & & & \\
          & & & & & & \\
         \midrule
         \midrule
         CIFAR-100, AlexNet & No-def & 0.9989 & 0.3054 & 0.00 & 0.00 & 0.00 & 93.96 & 171.95 & \textbf{357.37}\\
        CIFAR-100, AlexNet & Adv-reg & 0.9822 & 0.3222 & 0.00 & 0.00 & 0.00 & 38.36 & 247.44 & \textbf{335.18} \\
        CIFAR-100, AlexNet & Mem-guard & 0.9999 & 0.3048 & 0.00 & 0.00 & 0.00 & 55.67 & \textbf{218.70} & OOT \\
        CIFAR-100, AlexNet & DMP & MNC & MNC & MNC & MNC & MNC & MNC & MNC & MNC \\
        CIFAR-100, AlexNet & Mixup+MMD & 0.5473 & 0.3018 & 0.00 & 0.10 & 0.13 & 6.10 & 103.38 & \textbf{112.11}  \\
        CIFAR-100, AlexNet & SELENA & 0.3373 & 0.2879 & 0.00 & 0.15 & 0.20 & 2.68 & 20.99 & \textbf{35.09}\\
        CIFAR-100, AlexNet & HAMP & 0.5959 & 0.3147 & 0.00 & 0.13 & 0.19 & 3.89 & 31.50 & \textbf{86.93} \\
        CIFAR-100, AlexNet & DP-SGD & 0.5029 & 0.2977 & 0.00 & 0.33 & 0.40 & 10.38 & 18.87 & \textbf{31.75} \\
        CIFAR-100, AlexNet & MIST+Mixup & 0.3535 & 0.3049 & 0.00 & 0.14 & 0.19 & 6.92 & \textbf{17.43} & 9.45\\
        \midrule
        CIFAR-100, ResNet18 & No-def & 0.9999 & 0.5301 & 0.00 & 0.00 & 0.00 & 55.40 & 104.52 & \textbf{354.18}\\
        CIFAR-100, ResNet18 & Adv-reg & 0.5491 & 0.4455 & 0.00 & 0.20 & 0.20 & 0.82 & 3.61 & 10.29 \\
        CIFAR-100, ResNet18 & Mem-guard & 0.9999 & 0.5298 & 0.00 & 0.00 & 0.00 & 15.53 & \textbf{96.82} & OOT \\
        CIFAR-100, ResNet18 & DMP & 0.1585 & 0.1490 & 0.00 & 0.28 & 0.39 & 29.43 & 1.06 & 2.64 \\
        CIFAR-100, ResNet18 & Mixup+MMD & 0.8883 & 0.5197 & 0.00 & 0.00 & 0.00 & 3.83 & 186.31 & \textbf{259.35}\\
        CIFAR-100, ResNet18 & SELENA & OOT & OOT & OOT & OOT & OOT & OOT & OOT & OOT \\
        CIFAR-100, ResNet18 & HAMP & 0.9851 & 0.5188 & 0.00 & 0.20 & 0.20 & 20.11 & 236.08 & \textbf{412.31}\\
        CIFAR-100, ResNet18 & DP-SGD & 0.9804 & 0.5212 & 0.00 & 0.00 & 0.00 & 11.35 & 66.32 & \textbf{300.40} \\
        CIFAR-100, ResNet18 & MIST+Mixup & 0.5389 & 0.5202 & 0.00 & 0.10 & 0.13 & 5.49 & 27.60 & \textbf{68.34} \\ 
        \midrule
        CIFAR-100, DenseNet & No-def & 1.0000 & 0.5211 & 0.00 & 0.00 & 0.00 & 20.21 & 120.10 & \textbf{395.27} \\
        CIFAR-100, DenseNet & Adv-reg & 0.2891 & 0.2858 & 0.00 & 0.00 & 0.00 & 1.03 & 0.48 & 0.89 \\
        CIFAR-100, DenseNet & Mem-guard & 1.0000 & 0.5209 & 0.00 & 0.00 & 0.00 & 1.48 & \textbf{42.07} & OOT \\
        CIFAR-100, DenseNet & DMP & 0.0682 & 0.0665 & 0.00 & 0.00 & 0.00 & 1.10 & 1.05 & 1.13 \\
        CIFAR-100, DenseNet & Mixup+MMD & 0.8286 & 0.5177 & 0.00 & 0.18 & 0.23 & 200.09 & 99.16 & \textbf{272.22} \\
        CIFAR-100, DenseNet & SELENA & OOT & OOT & OOT & OOT & OOT & OOT & OOT  & OOT \\
        CIFAR-100, DenseNet & HAMP & 0.8397 & 0.5184 & 0.00 & 0.10 & 0.10 & 16.48 & 34.73 & \textbf{40.07} \\
        CIFAR-100, DenseNet & DP-SGD & 0.8350 & 0.5103 & 0.00 & 0.15 & 0.15 & 0.59 & 21.49 & \textbf{35.55} \\
        CIFAR-100, DenseNet & MIST+Mixup & 0.5577 & 0.5188 & 0.00 & 0.20 & 0.20 & 1.20 & \textbf{17.47} & 12.10 \\
        \bottomrule
    \end{tabular}
    }
    \caption{Defense evaluation CIFAR-100. OOT stands for out-of-time. The SELENA defense cannot be evaluated on ResNet18 and DenseNet because this defense requires more than half an year to run the experiments. The Mem-guard defense cannot be evaluated against CANARY attack also because of time constraint. MNC stands for Model Not Converging (training failure). DMP defense and Adv-reg defense will sometimes result in extremely low testing accuracy and thus the attack evaluation against these defenses is not meaningful. Lower PLR while preserving test accuracy is better.}
    
    \label{tab:full_def_eval_tpr_cifar100}
    \vspace{0.2cm}
\end{table*}

\begin{table*}[!t]
    \centering
    \begin{tabular}{lrrrrrrrrrr}
        \toprule
        \multirow{8}{3em}{\centering Dataset-Model} & \multirow{8}{3em}{\centering Defense} & \multirow{8}{3em}{\centering Train Acc.} & \multirow{8}{3em}{\centering Test Acc.}  & \multirow{8}{4em}{ \centering random-perturbation Attack PLR$\downarrow$\\ (FPR@ $0.1\%$) ~\cite{jayaraman2021revisiting}}  & \multirow{8}{3em}{ \centering LOSS Attack PLR$\downarrow$\\ (FPR@ $0.1\%$) ~\cite{yeom2018privacy}} & \multirow{8}{3em}{ \centering modified-entropy Attack PLR$\downarrow$\\ (FPR@ $0.1\%$) ~\cite{song2019membership}}  & \multirow{8}{3em}{ \centering class-NN Attack PLR$\downarrow$\\ (FPR@ $0.1\%$) ~\cite{shokri2017membership}} & \multirow{8}{3em}{\centering LIRA Attack \\ PLR$\downarrow$\\ (FPR@ $0.1\%$)~\cite{carlini2022membership}}  & \multirow{8}{3em}{\centering CANARY Attack PLR$\downarrow$ \\ (FPR@ $0.1\%$)~\cite{wen2022canary}} \\
          & & & & & & \\
          & & & & & & \\
          & & & & & & \\
          & & & & & & \\
          & & & & & & \\
          & & & & & & \\
          & & & & & & \\
         \midrule
         \midrule
        Texas & No-def &  0.6955 & 0.5679 & 0.00 & 0.22 & 1.05 & \textbf{12.64} & 1.31 & N/A \\
        Texas & Adv-reg & 0.3205 & 0.3081 & 0.00 & 0.00 & 0.00 & 4.76 & 1.23 & N/A \\
        Texas & Mem-guard & 0.6955 & 0.5689 & 0.00 & 0.17 & 0.28 & \textbf{11.22} & 1.30 &  N/A \\ 
        Texas & DMP & MNC & MNC & MNC & MNC & MNC & MNC & MNC &  MNC \\
        Texas & MMD & 0.6397 & 0.5598 & 0.00 & 0.23 & 0.23 & \textbf{8.33} & 1.63 & N/A \\ 
        Texas & SELENA & 0.5914 & 0.5366  & 0.00 & 0.10 & 0.15 & \textbf{8.14} & 1.48 & N/A \\ 
        Texas & HAMP & 0.5346 & 0.5190  & 0.00 & 0.25 & 0.35 & \textbf{8.01} & 0.79 & N/A \\
        Texas & DP-SGD & 0.6984 & 0.5682  & 0.00 & 0.03 & 0.03 & \textbf{9.80} & 1.18 & N/A \\
        Texas& MIST & 0.6396 & 0.5638  & 0.00 & 0.23 & 0.39 & \textbf{8.01} & 1.29 & N/A \\ 
        
        \midrule
        Purchase & No-def & 0.9998 & 0.9017 & 0.00 & 0.03 & 0.10 & 1.07 & \textbf{1.55} & N/A \\
        Purchase & Adv-reg & 0.3692 & 0.3636 & 0.00 & 0.93 & 1.01 & \textbf{36.68} & 0.91 &  N/A \\
        Purchase & Mem-guard & 0.9999 & 0.9017 & 0.00 & 0.00 & 0.00 & \textbf{1.58} & 0.96 &  N/A \\
        Purchase & DMP & MNC & MNC & MNC & MNC & MNC & MNC & MNC &  MNC \\
        Purchase & MMD & 0.9897 & 0.9041  & 0.00 & 0.00 & 0.00 & 1.22 & \textbf{1.37} & N/A \\
        Purchase  & SELENA & 0.9234 & 0.8717 & 0.00 & 0.50 & 0.60 & 1.22 & \textbf{1.39} & N/A \\
        Purchase & HAMP & 0.8926 & 0.8353  & 0.00 & 0.99 & 1.01 & \textbf{4.08} & 1.08 & N/A \\
        Purchase & DP-SGD & 0.9893 & 0.8995  & 0.00 & 0.00 & 0.00 & \textbf{1.29} & 1.19 & N/A \\
        Purchase & MIST & 0.9789 & 0.8944  & 0.00 & 0.05 & 0.05 & \textbf{1.08} & 0.67 & N/A \\
        \midrule
        Location & No-def & 1.0000 & 0.7213  & 0.00 & 0.00 & 0.00 & \textbf{2.89} & 1.19 & N/A \\
        Location & Adv-reg & 0.9447 & 0.6410 & 0.00 & 0.06 & 0.13 & \textbf{2.87} & 1.20 &  N/A \\
        Location & Mem-guard & 1.0000 & 0.7143 & 0.00 & 0.00 & 0.00 & \textbf{1.36} & 1.15 &  N/A \\
        Location & DMP & MNC & MNC & MNC & MNC & MNC & MNC & MNC &  MNC \\
        Location & MMD & 1.0000 & 0.6901  & 0.00 & 0.00 & 0.00 & \textbf{17.75} & 1.37 & N/A \\
        Location & SELENA & 0.7307 & 0.6647  & 0.00 & 0.00 & 0.00 & \textbf{4.81} & 1.71 & N/A \\
        Location & HAMP & 0.9885 & 0.7115  & 0.00 & 0.00 & 0.00 & 4.66 & \textbf{21.56} & N/A \\
        Location & DP-SGD & 1.0000 & 0.7113  & 0.00 & 0.00 & 0.00 & \textbf{2.78} & 1.10 & N/A \\
        Location & MIST & 0.9477 & 0.7159  & 0.00 & 0.00 & 0.00 & 0.71 & \textbf{1.15} & N/A \\
        \bottomrule
    \end{tabular}
    \caption{Our proposed defense evaluation on binary feature datasets. OOT stands for out-of-time. N/A stands for not applicable. The CANARY attack is not applicable to PURCHASE, TEXAS and LOCATION dataset because these three datasets only contain binary features. MNC stands for Model Not Converging (training failure). DMP defense and Adv-reg defense will sometimes result in extremely low testing accuracy and thus the attack evaluation against these defenses is not meaningful. Lower PLR while preserving test accuracy is better.}
    \label{tab:full_def_eval_tpr_binary}
\end{table*}

\begin{table*}
    \centering
    \begin{tabular}{lrrrrrrrrrr}
        \toprule
        \multirow{8}{3em}{\centering Dataset-Model} & \multirow{8}{3em}{\centering Defense} & \multirow{8}{3em}{\centering Train Acc.} & \multirow{8}{3em}{\centering Test Acc.}  & \multirow{8}{4em}{ \centering random-perturbation Attack AUC ~\cite{jayaraman2021revisiting}}  & \multirow{8}{3em}{ \centering LOSS Attack AUC ~\cite{yeom2018privacy}} & \multirow{8}{3em}{ \centering modified-entropy Attack AUC ~\cite{song2019membership}}  & \multirow{8}{3em}{ \centering class-NN Attack AUC ~\cite{shokri2017membership}} & \multirow{8}{3em}{\centering LIRA Attack \\ AUC ~\cite{carlini2022membership}}  & \multirow{8}{3em}{\centering CANARY Attack AUC ~\cite{wen2022canary}} \\
         & & & & & & \\
          & & & & & & \\
          & & & & & & \\
          & & & & & & \\
          & & & & & & \\
          & & & & & & \\
         \midrule
         \midrule
        CIFAR-10, AlexNet & No-def & 0.9989 & 0.6960 & 0.5000 & 0.7075 & 0.7101 & 0.7221 & 0.7598 & 0.8803 \\
        CIFAR-10, AlexNet & Adv-reg & 0.9980 & 0.7103 & 0.5000 & 0.6558 & 0.6694 & 0.6728 & 0.7454 & 0.8256 \\
        CIFAR-10, AlexNet & Mem-guard & 0.9930 & 0.7010 & 0.5000 & 0.6225 & 0.6275 & 0.6396 & 0.7259 & OOT \\
        CIFAR-10, AlexNet & DMP & 0.4141 & 0.4043 & 0.5010 & 0.5077 & 0.5088 & 0.5044 & 0.5067 & 0.5129 \\
        CIFAR-10, AlexNet & Mixup+MMD & 0.8805 & 0.7012 & 0.5000 & 0.6322 & 0.6382 & 0.6470 & 0.7014 & 0.8050 \\
        CIFAR-10, AlexNet & SELENA & 0.7250 & 0.6710 & 0.5000 & 0.5488 & 0.5483 & 0.5038 & 0.6338 & 0.6415 \\
        CIFAR-10, AlexNet & HAMP  & 0.8615 & 0.7106 & 0.5000 & 0.5895 & 0.5900 & 0.5854 & 0.6475 & 0.7559  \\
        CIFAR-10, AlexNet & DP-SGD & 0.9881 & 0.6925 & 0.5001 & 0.7032 & 0.7138 & 0.7160 & 0.7634 & 0.8687 \\
        CIFAR-10, AlexNet &MIST+Mixup& 0.7589  & 0.7006 & 0.5000 & 0.5838 & 0.5830 & 0.5878 & 0.6295 & 0.7237 \\
        \midrule
        CIFAR-10, ResNet18 & No-def & 0.9991 & 0.8631 & 0.5000 & 0.6144 & 0.6133 & 0.6159 & 0.6684 & 0.7775  \\
        CIFAR-10, ResNet18 & Adv-reg & MNC & MNC & MNC & MNC & MNC & MNC & MNC & MNC\\
        CIFAR-10, ResNet18 & Mem-guard & 0.9999 & 0.8630 & 0.5000 & 0.5830 & 0.5820 & 0.5691 & 0.6157 & OOT \\
        CIFAR-10, ResNet18 & DMP & 0.6973 & 0.6745 & 0.5000 & 0.5151 & 0.5167 & 0.5161 & 0.5084 & 0.5199 \\
        CIFAR-10, ResNet18 & Mixup+MMD & 0.9630 & 0.8685 & 0.5000 & 0.5770 & 0.5780 & 0.6221 & 0.6122 & 0.7640  \\
        CIFAR-10, ResNet18 & SELENA & OOT & OOT & OOT & OOT & OOT & OOT & OOT & OOT  \\
        CIFAR-10, ResNet18 & HAMP & 0.9957 & 0.8651 & 0.5000 & 0.6290 & 0.6293 & 0.6350 & 0.6830 & 0.7890\\
        CIFAR-10, ResNet18 & DP-SGD & 0.9797 & 0.8719 & 0.5000 & 0.5749 & 0.5788 & 0.6203 & 0.6190 & 0.6879 \\
        CIFAR-10, ResNet18 &MIST+Mixup& 0.8802 & 0.8633& 0.5000 & 0.5623 & 0.5666 & 0.5690 & 0.5795 & 0.6516  \\
        \midrule
        CIFAR-10, DenseNet & No-def & 0.9995 & 0.8788  & 0.5000 & 0.6208 & 0.6201 & 0.6178 & 0.6932 & 0.7575 \\
        CIFAR-10, DenseNet & Adv-reg & 0.2891 & 0.2841 & 0.5000 & 0.5037 & 0.5040 & 0.5079 & 0.5028 & 0.5088 \\
        CIFAR-10, DenseNet & Mem-guard & 0.9999 & 0.8788 & 0.5000 & 0.5736 & 0.5739 & 0.5760 & 0.6375 & OOT \\
        CIFAR-10, DenseNet & DMP & 0.4845 & 0.4785 & 0.5000 & 0.5031 & 0.5047 & 0.5020 & 0.5040 & 0.5091\\
        CIFAR-10, DenseNet & Mixup+MMD & 0.9577 & 0.8671 & 0.5000 & 0.5851 & 0.5890 & 0.5919 & 0.6744 & 0.7434 \\
        CIFAR-10, DenseNet & SELENA & OOT & OOT & OOT & OOT & OOT & OOT & OOT & OOT  \\
        CIFAR-10, DenseNet & HAMP & 0.9639 & 0.8780 & 0.5000 & 0.5680 & 0.5701 & 0.5893 & 0.6397 & 0.6918\\
        CIFAR-10, DenseNet & DP-SGD & 0.9898 & 0.8644 & 0.5000 & 0.5848 & 0.5830 & 0.5818 & 0.6319 & 0.6867 \\
        CIFAR-10, DenseNet &MIST+Mixup& 0.9097 & 0.8814 & 0.5000 & 0.5758 & 0.5799 & 0.5783 & 0.5991 & 0.6676  \\
        \bottomrule
    \end{tabular}
    \vspace{0.3cm}
    \caption{Our proposed defense evaluation on CIFAR-10. OOT stands for out-of-time. The SELENA defense cannot be evaluated on ResNet18 and DenseNet because this defense requires more than half an year to run the experiments. DMP defense and Adv-reg defense will sometimes result in extremely low testing accuracy and thus the attack evaluation against these defenses is not meaningful. MNC stands for Model Not Converging (training failure).}
    \label{tab:full_def_eval_auc_cifar10}
\end{table*}

\begin{table*}
    \centering
    \resizebox{1.02\textwidth}{!}{
    \begin{tabular}{lrrrrrrrrrr}
        \toprule
        \multirow{8}{3em}{\centering Dataset-Model} & \multirow{8}{3em}{\centering Defense} & \multirow{8}{3em}{\centering Train Acc.} & \multirow{8}{3em}{\centering Test Acc.}  & \multirow{8}{4em}{ \centering random-perturbation Attack AUC ~\cite{jayaraman2021revisiting}}  & \multirow{8}{3em}{ \centering LOSS Attack AUC ~\cite{yeom2018privacy}} & \multirow{8}{3em}{ \centering modified-entropy Attack AUC ~\cite{song2019membership}}  & \multirow{8}{3em}{ \centering class-NN Attack AUC ~\cite{shokri2017membership}} & \multirow{8}{3em}{\centering LIRA Attack \\ AUC ~\cite{carlini2022membership}}  & \multirow{8}{3em}{\centering CANARY Attack AUC ~\cite{wen2022canary}} \\
         & & & & & & \\
          & & & & & & \\
          & & & & & & \\
          & & & & & & \\
          & & & & & & \\
          & & & & & & \\
         \midrule
         \midrule
         CIFAR-100, AlexNet & No-def & 0.9989 & 0.3054 & 0.5000 & 0.8899 & 0.8920 & 0.9002 & 0.9242 & 0.9836 \\
        CIFAR-100, AlexNet & Adv-reg & 0.9822 & 0.3222 & 0.5000 & 0.8888 & 0.9010 & 0.9023 & 0.9482 & 0.9834 \\
        CIFAR-100, AlexNet & Mem-guard & 0.9999 & 0.3048 & 0.5000 & 0.8849 & 0.8890 & 0.9155 & 0.9593 & OOT\\
        CIFAR-100, AlexNet & DMP & MNC & MNC & MNC & MNC & MNC & MNC & MNC & MNC \\
        CIFAR-100, AlexNet & Mixup+MMD & 0.5473 & 0.3018 & 0.5000 & 0.7249 & 0.7299 & 0.7239 & 0.8089 & 0.9396  \\
        CIFAR-100, AlexNet & SELENA & 0.3373 & 0.2879 & 0.5000 & 0.5917 & 0.5901 & 0.6014 & 0.6644 & 0.7142 \\
        CIFAR-100, AlexNet & HAMP & 0.5959 & 0.3147 & 0.5000 & 0.6990 & 0.7120 & 0.7046 & 0.8120 & 0.8604 \\
        CIFAR-100, AlexNet & DP-SGD & 0.5029 & 0.2977 & 0.5002 & 0.6637 & 0.6710 & 0.6739 & 0.7014 & 0.8199  \\
        CIFAR-100, AlexNet &MIST+Mixup& 0.3535 & 0.3049 & 0.5000 & 0.5619 & 0.5693 & 0.5782 & 0.6412 & 0.7714 \\
        \midrule
        CIFAR-100, ResNet18 & No-def & 0.9999 & 0.5301 & 0.5000 & 0.8885 & 0.8899 & 0.9012 & 0.9270 & 0.9695 \\
        CIFAR-100, ResNet18 & Adv-reg & 0.5491 & 0.4455 & 0.5000 & 0.5830 & 0.5830 & 0.5719 & 0.5721 & 0.6320 \\
        CIFAR-100, ResNet18 & Mem-guard & 0.9999 & 0.5298 & 0.5000 & 0.7529 & 0.7623 & 0.7727 & 0.8669 & OOT\\
        CIFAR-100, ResNet18 & DMP & 0.1585 & 0.1490 & 0.5000 & 0.5113 & 0.5122 & 0.6521 & 0.5043 & 0.5139\\
        CIFAR-100, ResNet18 & Mixup+MMD & 0.8883 & 0.5197 & 0.5000 & 0.7845 & 0.7855 & 0.7910 & 0.9028 & 0.9288 \\
        CIFAR-100, ResNet18 & SELENA & OOT & OOT & OOT & OOT & OOT & OOT & OOT & OOT \\
        CIFAR-100, ResNet18 & HAMP & 0.9851 & 0.5188 & 0.5000 & 0.8308 & 0.8281 & 0.8502 & 0.9391 & 0.9710 \\
        CIFAR-100, ResNet18 & DP-SGD & 0.9804 & 0.5212 & 0.5000 & 0.7793 & 0.7801 & 0.7905 & 0.8552 & 0.9384 \\
        CIFAR-100, ResNet18 &MIST+Mixup& 0.5389 & 0.5202 & 0.5000 & 0.5877 & 0.5937 & 0.6394 & 0.6826 & 0.8288 \\ 
        \midrule
        CIFAR-100, DenseNet & No-def & 1.0000 & 0.5211 & 0.5000 & 0.8617 & 0.8610 & 0.8649 & 0.9270 & 0.9615 \\
        CIFAR-100, DenseNet & Adv-reg & 0.2891 & 0.2858 & 0.5000 & 0.5038 & 0.5047 & 0.5019 & 0.5060 & 0.5070 \\
        CIFAR-100, DenseNet & Mem-guard & 1.0000 & 0.5209 & 0.5000 & 0.7588 & 0.7601 & 0.7748 & 0.8469 & OOT \\
        CIFAR-100, DenseNet & DMP & 0.0682 & 0.0665  & 0.5000 & 0.5031 & 0.5048 & 0.5081 & 0.5093 & 0.5101 \\
        CIFAR-100, DenseNet & Mixup+MMD & 0.8286 & 0.5177 & 0.5000 & 0.8015 & 0.8103 & 0.8292 & 0.8385 & 0.9259 \\
        CIFAR-100, DenseNet & SELENA & OOT & OOT & OOT & OOT & OOT & OOT & OOT  & OOT \\
        CIFAR-100, DenseNet & HAMP & 0.8397 & 0.5184 & 0.5000 & 0.6864 & 0.6890 & 0.7564 & 0.7948 & 0.8257\\
        CIFAR-100, DenseNet & DP-SGD & 0.8350 & 0.5103 & 0.5000 & 0.6758 & 0.6794 & 0.6904 & 0.7418 & 0.7899 \\
        CIFAR-100, DenseNet & MIST+Mixup& 0.5577 & 0.5188 & 0.5000 & 0.5715 & 0.5839 & 0.5913 & 0.6868 & 0.7469 \\
        \bottomrule
    \end{tabular}
    }
    \vspace{0.3cm}
    \caption{Our proposed defense evaluation on CIFAR-100. OOT stands for out-of-time. The SELENA defense cannot be evaluated on ResNet18 and DenseNet because this defense requires more than half an year to run the experiments. DMP defense and Adv-reg defense will sometimes result in extremely low testing accuracy and thus the attack evaluation against these defenses is not meaningful. MNC stands for Model Not Converging (training failure).}
    \label{tab:full_def_eval_auc_cifar100}
\end{table*}

\begin{table*}
    \centering
    \begin{tabular}{lrrrrrrrrrr}
        \toprule
        \multirow{8}{3em}{\centering Dataset-Model} & \multirow{8}{3em}{\centering Defense} & \multirow{8}{3em}{\centering Train Acc.} & \multirow{8}{3em}{\centering Test Acc.}  & \multirow{8}{4em}{ \centering random-perturbation Attack AUC ~\cite{jayaraman2021revisiting}}  & \multirow{8}{3em}{ \centering LOSS Attack AUC ~\cite{yeom2018privacy}} & \multirow{8}{3em}{ \centering modified-entropy Attack AUC ~\cite{song2019membership}}  & \multirow{8}{3em}{ \centering class-NN Attack AUC ~\cite{shokri2017membership}} & \multirow{8}{3em}{\centering LIRA Attack \\ AUC ~\cite{carlini2022membership}}  & \multirow{8}{3em}{\centering CANARY Attack AUC ~\cite{wen2022canary}} \\
         & & & & & & \\
          & & & & & & \\
          & & & & & & \\
          & & & & & & \\
          & & & & & & \\
          & & & & & & \\
         \midrule
         \midrule
         
        Texas & No-def &  0.6955 & 0.5679 & 0.5000 & 0.6117 & 0.6203 & 0.6840 & 0.6211 & N/A \\
        Texas & Adv-reg & 0.3205 & 0.3081 & 0.5000 & 0.5136 & 0.5206 & 0.8450 & 0.5026 & N/A \\
        Texas & Mem-guard  & 0.6955 & 0.5689 & 0.5000 & 0.6186 & 0.6199 & 0.6735 & 0.5102 & N/A\\
        Texas & DMP & MNC & MNC & MNC & MNC & MNC & MNC & MNC & MNC \\
        Texas & MMD & 0.6397 & 0.5598 & 0.5000 & 0.5991 & 0.6001 & 0.6533 & 0.5545 & N/A \\
        Texas & SELENA & 0.5914 & 0.5366  & 0.5000 & 0.5498 & 0.5499 & 0.6344 & 0.5479 & N/A \\
        Texas & HAMP & 0.5346 & 0.5190  & 0.5010 & 0.5066 & 0.5109 & 0.7073 & 0.5019 & N/A \\
        Texas & DP-SGD & 0.6892 & 0.5596  & 0.5000 & 0.6105 & 0.6115 & 0.6795 & 0.5789 & N/A \\
        Texas& \EiSL & 0.6396 & 0.5638 & 0.5000 & 0.5800 & 0.5830 & 0.6379 & 0.5574 & N/A \\
        
        \midrule
        Purchase & No-def & 0.9998 & 0.9017  & 0.5020 & 0.6171 & 0.6210 & 0.6351 & 0.6689 & N/A \\
        Purchase & Adv-reg & 0.3692 & 0.3636 & 0.5000 & 0.5070 & 0.5080 & 0.5058 & 0.5004 & N/A \\
        Purchase & Mem-guard & 0.9999 & 0.9017 & 0.5000 & 0.5287 & 0.5290 & 0.5298 & 0.5080 & N/A\\
        Purchase & DMP & MNC & MNC & MNC & MNC & MNC & MNC & MNC & MNC \\
        Purchase & MMD & 0.9897 & 0.9041  & 0.5000 & 0.6052 & 0.6039 & 0.6300 & 0.6564 & N/A \\
        Purchase  & SELENA & 0.9234 & 0.8717 & 0.5000 & 0.5460 & 0.5480 & 0.5320 & 0.6592 & N/A \\
        Purchase & HAMP & 0.8926 & 0.8343  & 0.5000 & 0.5053 & 0.5038 & 0.5271 & 0.5004 & N/A \\
        Purchase & DP-SGD & 0.9893 & 0.8995  & 0.5000 & 0.6233 & 0.6310 & 0.6422 & 0.6732 & N/A \\
        Purchase & \EiSL & 0.9789 & 0.8944  & 0.5000 & 0.5975 & 0.5979 & 0.6150 & 0.5938  & N/A \\
        \midrule
        Location & No-def & 1.0000 & 0.7213  & 0.5002 & 0.7753 & 0.7758 & 0.7688 & 0.9258 & N/A \\
        Location & Adv-reg & 0.9447 & 0.6410 & 0.5010 & 0.6489 & 0.6481 & 0.6843 & 0.5708 & N/A \\
        Location & Mem-guard & 1.0000 & 0.7143 & 0.5000 & 0.7716 & 0.7720 & 0.7475 & 0.7790 & N/A \\
        Location & DMP & MNC & MNC & MNC & MNC & MNC & MNC & MNC & MNC \\
        Location & MMD & 1.0000 & 0.6901  & 0.5000 & 0.6270 & 0.6239 & 0.8094 & 0.6139 & N/A \\
        Location & SELENA & 0.7307 & 0.6647  & 0.5000 & 0.5116 & 0.5122 & 0.5079 & 0.5419 & N/A \\
        Location & HAMP & 0.9885 & 0.7115 & 0.5109 & 0.6868 & 0.6910 & 0.6835 & 0.8959 & N/A \\
        Location & DP-SGD & 1.0000 & 0.7113  & 0.5028 & 0.7827 & 0.7837 & 0.7797 & 0.8831 & N/A \\
        Location & \EiSL & 0.9560 & 0.7159  & 0.5001 & 0.7120 & 0.7111 & 0.7583 & 0.7255 & N/A \\
        \bottomrule
    \end{tabular}
    \vspace{0.3cm}
    \caption{Our proposed defense evaluation on binary feature datasets. OOT stands for out-of-time. N/A stands for not applicable. The CANARY attack is not applicable to PURCHASE, TEXAS and LOCATION dataset because these three datasets only contain binary features. MNC stands for Model Not Converging (training failure). DMP defense and Adv-reg defense will sometimes result in extremely low testing accuracy and thus the attack evaluation against these defenses is not meaningful.}
    \label{tab:full_def_eval_auc_binary}
\end{table*}


\begin{table*}[!t]
    \centering
    \begin{tabular}{lrrrrrrrrrr}
        \toprule
        \multirow{8}{3em}{\centering Dataset-Model} & \multirow{8}{3em}{\centering Defense} & \multirow{8}{3em}{\centering Train Acc.} & \multirow{8}{3em}{\centering Test Acc.}  & \multirow{8}{4em}{ \centering random-perturbation Attack PLR$\downarrow$\\ (FPR @ $1\%$) ~\cite{jayaraman2021revisiting}}  & \multirow{8}{3em}{ \centering LOSS Attack PLR$\downarrow$\\ (FPR @ $1\%$) ~\cite{yeom2018privacy}} & \multirow{8}{3em}{ \centering modified-entropy Attack PLR$\downarrow$\\ (FPR @ $1\%$) ~\cite{song2019membership}}  & \multirow{8}{3em}{ \centering class-NN Attack PLR$\downarrow$\\ (FPR @ $1\%$) ~\cite{shokri2017membership}} & \multirow{8}{3em}{\centering LIRA Attack \\ PLR$\downarrow$\\ (FPR @ $1\%$)~\cite{carlini2022membership}}  & \multirow{8}{3em}{\centering CANARY Attack PLR$\downarrow$ \\ (FPR @ $1\%$)~\cite{wen2022canary}} \\
         & & & & & & \\
          & & & & & & \\
          & & & & & & \\
          & & & & & & \\
          & & & & & & \\
          & & & & & & \\
          & & & & & & \\
          & & & & & & \\
         \midrule
         \midrule
        CIFAR-10, AlexNet & No-def & 0.9989 & 0.6960 & 0.00 & 0.00 & 0.00 & 0.47 & 18.78 & \textbf{35.69} \\
        CIFAR-10, AlexNet & Adv-reg & 0.9781 & 0.7103 & 0.00 & 0.00 & 0.00 & 3.03 & \textbf{43.29} & 25.66 \\
        CIFAR-10, AlexNet & Mem-guard & 0.9930 & 0.7010 & 0.00 & 0.15 & 0.15 & 0.47 & \textbf{16.64} & OOT \\
        CIFAR-10, AlexNet & DMP & 0.4141 & 0.4043 & 0.00 & 0.12 & 0.58 & 1.17 & 1.33 & 0.99 \\
        CIFAR-10, AlexNet & Mixup+MMD & 0.8805 & 0.7012 & 0.00 & 0.10 & 0.12 & 1.26 & 9.79 & \textbf{19.70}\\
        CIFAR-10, AlexNet & SELENA & 0.7250 & 0.6710 & 0.00 & 0.22 & 0.30 & 1.37 & 4.88 & \textbf{6.59} \\
        CIFAR-10, AlexNet & HAMP & 0.8615 & 0.7106 & 0.00 & 0.12 & 0.12 & 1.13 & 5.42 & \textbf{11.67} \\
        CIFAR-10, AlexNet & DP-SGD & 0.9881 & 0.6925 & 0.00 & 0.10 & 0.10 & 0.45 & 5.63 & \textbf{29.30}\\
        CIFAR-10, AlexNet & MIST+Mixup & 0.7589  & 0.7006 & 0.00 & 0.10 & 0.10 & 3.03 & 5.16 & \textbf{6.44}\\
        \midrule
        CIFAR-10, ResNet18 & No-def & 0.9991 & 0.8631 & 0.00 & 0.00 & 0.00 & 0.00 & 2.09 & \textbf{19.20} \\
        CIFAR-10, ResNet18 & Adv-reg & MNC & MNC & MNC & MNC & MNC & MNC & MNC & MNC\\
        CIFAR-10, ResNet18 & Mem-guard & 0.9999 & 0.8630 & 0.00 & 0.00 & 0.00 & 1.20 & \textbf{2.40} & OOT \\
        CIFAR-10, ResNet18 & DMP & 0.6973 & 0.6745 & 0.00 & 0.19 & 0.56 & 1.02 & 1.17 & 2.48 \\
        CIFAR-10, ResNet18 & Mixup+MMD & 0.9630 & 0.8685 & 0.00 & 0.05 & 0.10 & 1.53 & 2.55 & \textbf{18.20} \\
        CIFAR-10, ResNet18 & SELENA & OOT & OOT & OOT & OOT & OOT & OOT & OOT & OOT \\
        CIFAR-10, ResNet18 & HAMP & 0.9957 & 0.8651 & 0.00 & 0.00 & 0.00 & 1.85 & 9.18 & \textbf{17.90} \\
        CIFAR-10, ResNet18 & DP-SGD & 0.9797 & 0.8719 & 0.00 & 0.00 & 0.00 & 0.00 & 13.32 & \textbf{11.60} \\
        CIFAR-10, ResNet18 & MIST+Mixup & 0.8802 & 0.8633 & 0.00 & 0.10 & 0.10 & 1.51 & 2.28 & \textbf{1.57} \\
        \midrule
        CIFAR-10, DenseNet & No-def & 0.9995 & 0.8788  & 0.00 & 0.00 & 0.00 & 1.20 & 3.54 & \textbf{18.43} \\
        CIFAR-10, DenseNet & Adv-reg & 0.2891 & 0.2841 & 0.00 & 0.00 & 0.00 & 1.30 & 3.99 & 1.02 \\
        CIFAR-10, DenseNet & Mem-guard & 0.9999 & 0.8779 & 0.00 & 0.00 & 0.00 & 1.32 & \textbf{3.48} & OOT \\
        CIFAR-10, DenseNet & DMP & 0.4845 & 0.4785 & 0.00 & 0.888 & 1.01 & 1.22 & 2.59 & 2.53 \\
        CIFAR-10, DenseNet & Mixup+MMD & 0.9577 & 0.8671 & 0.00 & 0.10 & 0.10 & 8.23 & 1.81 & \textbf{15.83}\\
        CIFAR-10, DenseNet & SELENA & OOT & OOT & OOT & OOT & OOT & OOT & OOT & OOT  \\
        CIFAR-10, DenseNet & HAMP & 0.9639 & 0.8780 & 0.00 & 0.10 & 0.19 & 1.31 & 2.54 & \textbf{8.53}\\
        CIFAR-10, DenseNet & DP-SGD & 0.9898 & 0.8644 & 0.00 & 0.00 & 0.00 & 0.00 & 1.86 & \textbf{10.59} \\
        CIFAR-10, DenseNet & MIST+Mixup & 0.9097 & 0.8814 & 0.00 & 0.10 & 0.15 & 1.55 & 1.13 & \textbf{8.16} \\
        \bottomrule
    \end{tabular}
    \caption{Defense evaluation on CIFAR-10 with FPR = $0.01$. OOT stands for out-of-time. The SELENA defense cannot be evaluated on ResNet18 and DenseNet because this defense requires more than half an year to run the experiments. The Mem-guard defense cannot be evaluated against CANARY attack also because of time constraint. MNC stands for Model Not Converging (training failure). DMP defense and Adv-reg defense will sometimes result in extremely low testing accuracy and thus the attack evaluation against these defenses is not meaningful. Lower PLR while preserving test accuracy is better.}
    \label{tab:full_def_eval_tpr_cifar10_fpr001}
\end{table*}

\begin{table*}[!t]
    \centering
    \resizebox{1.02\textwidth}{!}{
    \begin{tabular}{lrrrrrrrrrr}
        \toprule
        \multirow{8}{7em}{\centering Dataset-Model} & \multirow{8}{3em}{\centering Defense} & \multirow{8}{3em}{\centering Train Acc.} & \multirow{8}{3em}{\centering Test Acc.}  & \multirow{8}{4em}{ \centering random-perturbation Attack PLR$\downarrow$\\ (FPR@ $1\%$) ~\cite{jayaraman2021revisiting}}  & \multirow{8}{3em}{ \centering LOSS Attack PLR$\downarrow$\\ (FPR@ $1\%$) ~\cite{yeom2018privacy}} & \multirow{8}{3em}{ \centering  modified-entropy Attack PLR$\downarrow$\\ (FPR@ $1\%$) ~\cite{song2019membership}}  & \multirow{8}{3em}{ \centering class-NN Attack PLR$\downarrow$\\ (FPR@ $1\%$) ~\cite{shokri2017membership}} & \multirow{8}{3em}{\centering LIRA Attack \\ PLR$\downarrow$\\ (FPR@ $1\%$)\cite{carlini2022membership}}  & \multirow{8}{3em}{\centering CANARY Attack PLR$\downarrow$ \\ (FPR@ $1\%$)\cite{wen2022canary}} \\
          & & & & & & \\
          & & & & & & \\
          & & & & & & \\
          & & & & & & \\
          & & & & & & \\
          & & & & & & \\
          & & & & & & \\
         \midrule
         \midrule
         CIFAR-100, AlexNet & No-def & 0.9989 & 0.3054 & 0.00 & 0.00 & 0.00 & 8.81 & 25.65 & \textbf{70.66}\\
        CIFAR-100, AlexNet & Adv-reg & 0.9822 & 0.3222 & 0.00 & 0.00 & 0.00 & 0.00 & 8.18 & \textbf{65.61} \\
        CIFAR-100, AlexNet & Mem-guard & 0.9999 & 0.3048 & 0.00 & 0.00 & 0.00 & 8.83 & \textbf{32.19} & OOT \\
        CIFAR-100, AlexNet & DMP & MNC & MNC & MNC & MNC & MNC & MNC & MNC & MNC \\
        CIFAR-100, AlexNet & Mixup+MMD & 0.5473 & 0.3018 & 0.00 & 0.10 & 0.13 & 7.69 & 23.13 & \textbf{28.89}  \\
        CIFAR-100, AlexNet & SELENA & 0.3373 & 0.2879 & 0.00 & 0.15 & 0.20 & 1.38  & 4.59 & \textbf{9.26}\\
        CIFAR-100, AlexNet & HAMP & 0.5959 & 0.3147 & 0.00 & 0.13 & 0.19 & 3.05 & 17.00 & \textbf{21.43} \\
        CIFAR-100, AlexNet & DP-SGD & 0.5029 & 0.2977 & 0.00 & 0.33 & 0.40 & 2.92 & 8.76 & \textbf{30.13} \\
        CIFAR-100, AlexNet & MIST+Mixup & 0.3535 & 0.3049 & 0.00 & 0.14 & 0.19 & 3.33 & 6.78 & \textbf{6.79} \\
        \midrule
        CIFAR-100, ResNet18 & No-def & 0.9999 & 0.5301 & 0.00 & 0.00 & 0.00 & 2.22 & 21.87 & \textbf{50.39}\\
        CIFAR-100, ResNet18 & Adv-reg & 0.5491 & 0.4455 & 0.00 & 0.20 & 0.20 & 0.82 & 3.61 & 10.29 \\
        CIFAR-100, ResNet18 & Mem-guard & 0.9999 & 0.5298 & 0.00 & 0.00 & 0.00 & 3.58 & \textbf{20.99} & OOT \\
        CIFAR-100, ResNet18 & DMP & 0.1585 & 0.1490 & 0.00 & 0.28 & 0.39 & 2.13 & 2.06 & 2.14 \\
        CIFAR-100, ResNet18 & Mixup+MMD & 0.8883 & 0.5197 & 0.00 & 0.00 & 0.00 & 7.43 & 29.48 & \textbf{74.89}\\
        CIFAR-100, ResNet18 & SELENA & OOT & OOT & OOT & OOT & OOT & OOT & OOT & OOT \\
        CIFAR-100, ResNet18 & HAMP & 0.9851 & 0.5188 & 0.00 & 0.20 & 0.20 & 7.46 & 42.98 & \textbf{45.39}\\
        CIFAR-100, ResNet18 & DP-SGD & 0.9804 & 0.5212 & 0.00 & 0.00 & 0.00 & 1.82 & 18.95 & \textbf{36.29} \\
        CIFAR-100, ResNet18 & MIST+Mixup & 0.5389 & 0.5202 & 0.00 & 0.10 & 0.13 & 4.97 & 12.19 & \textbf{21.39} \\ 
        \midrule
        CIFAR-100, DenseNet & No-def & 1.0000 & 0.5211 & 0.00 & 0.00 & 0.00 & 6.46 & 11.91 & \textbf{58.22} \\
        CIFAR-100, DenseNet & Adv-reg & 0.2891 & 0.2858 & 0.00 & 0.00 & 0.00 & 1.13 & 1.18 & 0.29 \\
        CIFAR-100, DenseNet & Mem-guard & 1.0000 & 0.5209 & 0.00 & 0.00 & 0.00 & 5.48 & \textbf{10.99} & OOT \\
        CIFAR-100, DenseNet & DMP & 0.0682 & 0.0665 & 0.00 & 0.00 & 0.00 & 1.10 & 1.05 & 1.13 \\
        CIFAR-100, DenseNet & Mixup+MMD & 0.8286 & 0.5177 & 0.00 & 0.18 & 0.23 & 25.40 & 1.38 & \textbf{41.30} \\
        CIFAR-100, DenseNet & SELENA & OOT & OOT & OOT & OOT & OOT & OOT & OOT  & OOT \\
        CIFAR-100, DenseNet & HAMP & 0.8397 & 0.5184 & 0.00 & 0.10 & 0.10 & 10.98 & 10.59 & \textbf{10.92} \\
        CIFAR-100, DenseNet & DP-SGD & 0.8350 & 0.5103 & 0.00 & 0.15 & 0.15 & 1.24 & 8.94 & \textbf{11.33} \\
        CIFAR-100, DenseNet & MIST+Mixup & 0.5577 & 0.5188 & 0.00 & 0.20 & 0.20 & 3.16 & 6.81 & \textbf{10.20} \\
        \bottomrule
    \end{tabular}
    }
    \caption{Defense evaluation CIFAR-100 with FPR = $0.01$. OOT stands for out-of-time. The SELENA defense cannot be evaluated on ResNet18 and DenseNet because this defense requires more than half an year to run the experiments. The Mem-guard defense cannot be evaluated against CANARY attack also because of time constraint. MNC stands for Model Not Converging (training failure). DMP defense and Adv-reg defense will sometimes result in extremely low testing accuracy and thus the attack evaluation against these defenses is not meaningful. Lower PLR while preserving test accuracy is better.}
    
    \label{tab:full_def_eval_tpr_cifar100_fpr001}
    \vspace{0.2cm}
\end{table*}

\begin{table*}[!t]
    \centering
    \begin{tabular}{lrrrrrrrrrr}
        \toprule
        \multirow{8}{3em}{\centering Dataset-Model} & \multirow{8}{3em}{\centering Defense} & \multirow{8}{3em}{\centering Train Acc.} & \multirow{8}{3em}{\centering Test Acc.}  & \multirow{8}{4em}{ \centering random-perturbation Attack PLR$\downarrow$\\ (FPR@ $1\%$) ~\cite{jayaraman2021revisiting}}  & \multirow{8}{3em}{ \centering LOSS Attack PLR$\downarrow$\\ (FPR@ $1\%$) ~\cite{yeom2018privacy}} & \multirow{8}{3em}{ \centering modified-entropy Attack PLR$\downarrow$\\ (FPR@ $1\%$) ~\cite{song2019membership}}  & \multirow{8}{3em}{ \centering class-NN Attack PLR$\downarrow$\\ (FPR@ $1\%$) ~\cite{shokri2017membership}} & \multirow{8}{3em}{\centering LIRA Attack \\ PLR$\downarrow$\\ (FPR@ $1\%$)~\cite{carlini2022membership}}  & \multirow{8}{3em}{\centering CANARY Attack PLR$\downarrow$ \\ (FPR@ $1\%$)~\cite{wen2022canary}} \\
          & & & & & & \\
          & & & & & & \\
          & & & & & & \\
          & & & & & & \\
          & & & & & & \\
          & & & & & & \\
          & & & & & & \\
         \midrule
         \midrule
        Texas & No-def &  0.6955 & 0.5679 & 0.00 & 0.22 & 1.05 & \textbf{3.88} & 1.25 & N/A \\
        Texas & Adv-reg & 0.3205 & 0.3081 & 0.00 & 0.00 & 0.00 & 3.28 & 1.18 & N/A \\
        Texas & Mem-guard & 0.6955 & 0.5689 & 0.00 & 0.17 & 0.28 & \textbf{4.69} & 1.30 &  N/A \\ 
        Texas & DMP & MNC & MNC & MNC & MNC & MNC & MNC & MNC &  MNC \\
        Texas & MMD & 0.6397 & 0.5598 & 0.00 & 0.23 & 0.23 & \textbf{2.69} & 1.19 & N/A \\ 
        Texas & SELENA & 0.5914 & 0.5366  & 0.00 & 0.10 & 0.15 & \textbf{3.01} & 1.13 & N/A \\ 
        Texas & HAMP & 0.5346 & 0.5190  & 0.00 & 0.25 & 0.35 & \textbf{4.08} & 1.03 & N/A \\
        Texas & DP-SGD & 0.6984 & 0.5682  & 0.00 & 0.03 & 0.03 & \textbf{3.88} & 1.38 & N/A \\
        Texas& MIST & 0.6396 & 0.5638  & 0.00 & 0.23 & 0.39 & \textbf{3.23} & 1.17 & N/A \\ 
        
        \midrule
        Purchase & No-def & 0.9998 & 0.9017 & 0.00 & 0.03 & 0.10 & 1.11 & \textbf{1.24} & N/A \\
        Purchase & Adv-reg & 0.3692 & 0.3636 & 0.00 & 0.93 & 1.01 & 0.95 & 1.13 &  N/A \\
        Purchase & Mem-guard & 0.9999 & 0.9017 & 0.00 & 0.00 & 0.00 & \textbf{1.33} & 1.10 &  N/A \\
        Purchase & DMP & MNC & MNC & MNC & MNC & MNC & MNC & MNC &  MNC \\
        Purchase & MMD & 0.9897 & 0.9041  & 0.00 & 0.00 & 0.00 & 0.94 & \textbf{1.13} & N/A \\
        Purchase  & SELENA & 0.9234 & 0.8717 & 0.00 & 0.50 & 0.60 & \textbf{1.30} & 1.11 & N/A \\
        Purchase & HAMP & 0.8926 & 0.8353  & 0.00 & 0.99 & 1.01 & \textbf{1.44} & 0.86 & N/A \\
        Purchase & DP-SGD & 0.9893 & 0.8995  & 0.00 & 0.00 & 0.00 & \textbf{1.37} & 0.82 & N/A \\
        Purchase & MIST & 0.9789 & 0.8944  & 0.00 & 0.05 & 0.05 & \textbf{1.17} & 0.67 & N/A \\
        \midrule
        Location & No-def & 1.0000 & 0.7213  & 0.00 & 0.00 & 0.00 & 1.50 & \textbf{13.19} & N/A \\
        Location & Adv-reg & 0.9447 & 0.6410 & 0.00 & 0.06 & 0.13 & 1.53 & \textbf{13.39} &  N/A \\
        Location & Mem-guard & 1.0000 & 0.7143 & 0.00 & 0.00 & 0.00 & \textbf{1.66} & 1.10 &  N/A \\
        Location & DMP & MNC & MNC & MNC & MNC & MNC & MNC & MNC &  MNC \\
        Location & MMD & 1.0000 & 0.6901  & 0.00 & 0.00 & 0.00 & \textbf{2.95} & 0.88 & N/A \\
        Location & SELENA & 0.7307 & 0.6647  & 0.00 & 0.00 & 0.00 & \textbf{1.30} & 1.03 & N/A \\
        Location & HAMP & 0.9885 & 0.7115  & 0.00 & 0.00 & 0.00 & 1.22 & \textbf{12.30} & N/A \\
        Location & DP-SGD & 1.0000 & 0.7113  & 0.00 & 0.00 & 0.00 & 1.33 & \textbf{3.03} & N/A \\
        Location & MIST & 0.9477 & 0.7159  & 0.00 & 0.00 & 0.00 & \textbf{1.55} & 1.27 & N/A \\
        \bottomrule
    \end{tabular}
    \caption{Our proposed defense evaluation on binary feature datasets with FPR = $0.01$. OOT stands for out-of-time. N/A stands for not applicable. The CANARY attack is not applicable to PURCHASE, TEXAS and LOCATION dataset because these three datasets only contain binary features. MNC stands for Model Not Converging (training failure). DMP defense and Adv-reg defense will sometimes result in extremely low testing accuracy and thus the attack evaluation against these defenses is not meaningful. Lower PLR while preserving test accuracy is better.}
    \label{tab:full_def_eval_tpr_binary_fpr001}
\end{table*}


\begin{table*}[!t]
    \centering
    \begin{tabular}{lrrrrrrrrrr}
        \toprule
        \multirow{8}{3em}{\centering Dataset-Model} & \multirow{8}{3em}{\centering Defense} & \multirow{8}{3em}{\centering Train Acc.} & \multirow{8}{3em}{\centering Test Acc.}  & \multirow{8}{4em}{ \centering random-perturbation Attack PLR$\downarrow$\\ (FPR @ $0.5\%$) ~\cite{jayaraman2021revisiting}}  & \multirow{8}{3em}{ \centering LOSS Attack PLR$\downarrow$\\ (FPR @ $0.5\%$) ~\cite{yeom2018privacy}} & \multirow{8}{3em}{ \centering modified-entropy Attack PLR$\downarrow$\\ (FPR @ $0.5\%$) ~\cite{song2019membership}}  & \multirow{8}{3em}{ \centering class-NN Attack PLR$\downarrow$\\ (FPR @ $0.5\%$) ~\cite{shokri2017membership}} & \multirow{8}{3em}{\centering LIRA Attack \\ PLR$\downarrow$\\ (FPR @ $0.5\%$)~\cite{carlini2022membership}}  & \multirow{8}{3em}{\centering CANARY Attack PLR$\downarrow$ \\ (FPR @ $0.5\%$)~\cite{wen2022canary}} \\
         & & & & & & \\
          & & & & & & \\
          & & & & & & \\
          & & & & & & \\
          & & & & & & \\
          & & & & & & \\
          & & & & & & \\
          & & & & & & \\
         \midrule
         \midrule
        CIFAR-10, AlexNet & No-def & 0.9989 & 0.6960 & 0.00 & 0.00 & 0.00 & 0.00 & 8.32 & \textbf{57.21} \\
        CIFAR-10, AlexNet & Adv-reg & 0.9781 & 0.7103 & 0.00 & 0.00 & 0.00 & 0.00 & 7.88 & \textbf{40.52} \\
        CIFAR-10, AlexNet & Mem-guard & 0.9930 & 0.7010 & 0.00 & 0.15 & 0.15 & 0.17 & \textbf{10.03} & OOT \\
        CIFAR-10, AlexNet & DMP & 0.4141 & 0.4043 & 0.00 & 0.38 & 0.50 & 1.17 & 1.03 & 0.91 \\
        CIFAR-10, AlexNet & Mixup+MMD & 0.8805 & 0.7012 & 0.00 & 0.10 & 0.12 & 1.49 & 13.22 & \textbf{31.20}\\
        CIFAR-10, AlexNet & SELENA & 0.7250 & 0.6710 & 0.00 & 0.22 & 0.30 & 1.49 & 6.38 & \textbf{7.90} \\
        CIFAR-10, AlexNet & HAMP & 0.8615 & 0.7106 & 0.00 & 0.12 & 0.12 & 1.15 & 5.26 & \textbf{14.40} \\
        CIFAR-10, AlexNet & DP-SGD & 0.9881 & 0.6925 & 0.00 & 0.10 & 0.10 & 0.00 & 8.05 & \textbf{43.62}\\
        CIFAR-10, AlexNet & MIST+Mixup & 0.7589  & 0.7006 & 0.00 & 0.10 & 0.10 & 3.21 & 6.54 & \textbf{9.62}\\
        \midrule
        CIFAR-10, ResNet18 & No-def & 0.9991 & 0.8631 & 0.00 & 0.00 & 0.00 & 0.00 & 3.04 & \textbf{28.01} \\
        CIFAR-10, ResNet18 & Adv-reg & MNC & MNC & MNC & MNC & MNC & MNC & MNC & MNC\\
        CIFAR-10, ResNet18 & Mem-guard & 0.9999 & 0.8630 & 0.00 & 0.00 & 0.00 & 0.00 & \textbf{3.33} & OOT \\
        CIFAR-10, ResNet18 & DMP & 0.6973 & 0.6745 & 0.00 & 0.49 & 0.59 & 1.69 & 1.77 & 4.88 \\
        CIFAR-10, ResNet18 & Mixup+MMD & 0.9630 & 0.8685 & 0.00 & 0.05 & 0.10 & 1.59 & 3.39 & \textbf{29.20} \\
        CIFAR-10, ResNet18 & SELENA & OOT & OOT & OOT & OOT & OOT & OOT & OOT & OOT  \\
        CIFAR-10, ResNet18 & HAMP & 0.9957 & 0.8651 & 0.00 & 0.00 & 0.00 & 2.02 & 13.11 & \textbf{30.80}\\
        CIFAR-10, ResNet18 & DP-SGD & 0.9797 & 0.8719 & 0.00 & 0.00 & 0.00 & 5.05 & 3.13 & \textbf{15.60} \\
        CIFAR-10, ResNet18 & MIST+Mixup & 0.8802 & 0.8633 & 0.00 & 0.10 & 0.10 & 1.53 & 2.31 & \textbf{2.62}\\
        \midrule
        CIFAR-10, DenseNet & No-def & 0.9995 & 0.8788  & 0.00 & 0.00 & 0.00 & 10.99 & 15.98 & \textbf{30.88} \\
        CIFAR-10, DenseNet & Adv-reg & 0.2891 & 0.2841 & 0.00 & 0.00 & 0.00 & 0.12 & 1.89 & 2.01 \\
        CIFAR-10, DenseNet & Mem-guard & 0.9999 & 0.8779 & 0.00 & 0.00 & 0.00 & 5.83 & \textbf{14.35} & OOT \\
        CIFAR-10, DenseNet & DMP & 0.4845 & 0.4785 & 0.00 & 0.38 & 1.17 & 2.28 & 3.41 & 3.59 \\
        CIFAR-10, DenseNet & Mixup+MMD & 0.9577 & 0.8671 & 0.00 & 0.10 & 0.10 & 13.04 & 14.11 & \textbf{25.80}\\
        CIFAR-10, DenseNet & SELENA & OOT & OOT & OOT & OOT & OOT & OOT & OOT & OOT  \\
        CIFAR-10, DenseNet & HAMP & 0.9639 & 0.8780 & 0.00 & 0.10 & 0.19 & 1.33 & 3.17 & \textbf{14.65}\\
        CIFAR-10, DenseNet & DP-SGD & 0.9898 & 0.8644 & 0.00 & 0.00 & 0.00 & 1.47 & 2.67 & \textbf{25.39} \\
        CIFAR-10, DenseNet & MIST+Mixup & 0.9097 & 0.8814 & 0.00 & 0.10 & 0.15 & 1.48 & 1.77 & \textbf{12.12} \\
        \bottomrule
    \end{tabular}
    \caption{Defense evaluation on CIFAR-10 with FPR = $0.005$. OOT stands for out-of-time. The SELENA defense cannot be evaluated on ResNet18 and DenseNet because this defense requires more than half an year to run the experiments. The Mem-guard defense cannot be evaluated against CANARY attack also because of time constraint. MNC stands for Model Not Converging (training failure). DMP defense and Adv-reg defense will sometimes result in extremely low testing accuracy and thus the attack evaluation against these defenses is not meaningful. Lower PLR while preserving test accuracy is better.}
    \label{tab:full_def_eval_tpr_cifar10_fpr0005}
\end{table*}

\begin{table*}[!t]
    \centering
    \resizebox{1.02\textwidth}{!}{
    \begin{tabular}{lrrrrrrrrrr}
        \toprule
        \multirow{8}{7em}{\centering Dataset-Model} & \multirow{8}{3em}{\centering Defense} & \multirow{8}{3em}{\centering Train Acc.} & \multirow{8}{3em}{\centering Test Acc.}  & \multirow{8}{4em}{ \centering random-perturbation Attack PLR$\downarrow$\\ (FPR@ $0.5\%$) ~\cite{jayaraman2021revisiting}}  & \multirow{8}{3em}{ \centering LOSS Attack PLR$\downarrow$\\ (FPR@ $0.5\%$) ~\cite{yeom2018privacy}} & \multirow{8}{3em}{ \centering  modified-entropy Attack PLR$\downarrow$\\ (FPR@ $0.5\%$) ~\cite{song2019membership}}  & \multirow{8}{3em}{ \centering class-NN Attack PLR$\downarrow$\\ (FPR@ $0.5\%$) ~\cite{shokri2017membership}} & \multirow{8}{3em}{\centering LIRA Attack \\ PLR$\downarrow$\\ (FPR@ $0.5\%$)\cite{carlini2022membership}}  & \multirow{8}{3em}{\centering CANARY Attack PLR$\downarrow$ \\ (FPR@ $0.5\%$)\cite{wen2022canary}} \\
          & & & & & & \\
          & & & & & & \\
          & & & & & & \\
          & & & & & & \\
          & & & & & & \\
          & & & & & & \\
          & & & & & & \\
         \midrule
         \midrule
         CIFAR-100, AlexNet & No-def & 0.9989 & 0.3054 & 0.00 & 0.00 & 0.00 & 7.49 & 48.31 & \textbf{103.60}\\
        CIFAR-100, AlexNet & Adv-reg & 0.9822 & 0.3222 & 0.00 & 0.00 & 0.00 & 6.94 & 45.23 & \textbf{116.31} \\
        CIFAR-100, AlexNet & Mem-guard & 0.9999 & 0.3048 & 0.00 & 0.00 & 0.00 & 7.40 & \textbf{56.60} & OOT \\
        CIFAR-100, AlexNet & DMP & MNC & MNC & MNC & MNC & MNC & MNC & MNC & MNC \\
        CIFAR-100, AlexNet & Mixup+MMD & 0.5473 & 0.3018 & 0.00 & 0.10 & 0.13 & 6.07 & 35.44 & \textbf{45.42}  \\
        CIFAR-100, AlexNet & SELENA & 0.3373 & 0.2879 & 0.00 & 0.15 & 0.20 & 1.76 & 7.49 & \textbf{13.20}\\
        CIFAR-100, AlexNet & HAMP & 0.5959 & 0.3147 & 0.00 & 0.13 & 0.19 & 3.06 & 24.39 & \textbf{29.84} \\
        CIFAR-100, AlexNet & DP-SGD & 0.5029 & 0.2977 & 0.00 & 0.33 & 0.40 & 3.17 & 11.43 & \textbf{38.40} \\
        CIFAR-100, AlexNet & MIST+Mixup & 0.3535 & 0.3049 & 0.00 & 0.14 & 0.19 & 3.45 & \textbf{8.86} & 8.63\\
        \midrule
        CIFAR-100, ResNet18 & No-def & 0.9999 & 0.5301 & 0.00 & 0.00 & 0.00 & 22.07 & 31.60 & \textbf{160.48}\\
        CIFAR-100, ResNet18 & Adv-reg & 0.5491 & 0.4455 & 0.00 & 0.20 & 0.20 & 5.82 & 7.62 & 31.49 \\
        CIFAR-100, ResNet18 & Mem-guard & 0.9999 & 0.5298 & 0.00 & 0.00 & 0.00 & 20.38 & \textbf{35.58} & OOT \\
        CIFAR-100, ResNet18 & DMP & 0.1585 & 0.1490 & 0.00 & 0.28 & 0.39 & 21.41 & 5.06 & 4.64 \\
        CIFAR-100, ResNet18 & Mixup+MMD & 0.8883 & 0.5197 & 0.00 & 0.00 & 0.00 & 7.44 & 45.91 & \textbf{131.10}\\
        CIFAR-100, ResNet18 & SELENA & OOT & OOT & OOT & OOT & OOT & OOT & OOT & OOT \\
        CIFAR-100, ResNet18 & HAMP & 0.9851 & 0.5188 & 0.00 & 0.20 & 0.20 & 5.33 & 70.59 & \textbf{110.31}\\
        CIFAR-100, ResNet18 & DP-SGD & 0.9804 & 0.5212 & 0.00 & 0.00 & 0.00 & 1.36 & 23.43 & \textbf{55.10} \\
        CIFAR-100, ResNet18 & MIST+Mixup & 0.5389 & 0.5202 & 0.00 & 0.10 & 0.13 & 11.49 & 16.89 & \textbf{51.78} \\ 
        \midrule
        CIFAR-100, DenseNet & No-def & 1.0000 & 0.5211 & 0.00 & 0.00 & 0.00 & 4.09 & 58.30 & \textbf{98.44} \\
        CIFAR-100, DenseNet & Adv-reg & 0.2891 & 0.2858 & 0.00 & 0.00 & 0.00 & 2.03 & 0.49 & 1.19 \\
        CIFAR-100, DenseNet & Mem-guard & 1.0000 & 0.5209 & 0.00 & 0.00 & 0.00 & 8.48 & \textbf{18.94} & OOT \\
        CIFAR-100, DenseNet & DMP & 0.0682 & 0.0665 & 0.00 & 0.00 & 0.00 & 1.30 & 1.25 & 1.33 \\
        CIFAR-100, DenseNet & Mixup+MMD & 0.8286 & 0.5177 & 0.00 & 0.18 & 0.23 & 46.94 & 14.33 & \textbf{71.32} \\
        CIFAR-100, DenseNet & SELENA & OOT & OOT & OOT & OOT & OOT & OOT & OOT  & OOT \\
        CIFAR-100, DenseNet & HAMP & 0.8397 & 0.5184 & 0.00 & 0.10 & 0.10 & 13.45 & 14.39 & \textbf{20.95} \\
        CIFAR-100, DenseNet & DP-SGD & 0.8350 & 0.5103 & 0.00 & 0.15 & 0.15 & 0.99 & 9.04 & \textbf{15.48} \\
        CIFAR-100, DenseNet & MIST+Mixup & 0.5577 & 0.5188 & 0.00 & 0.20 & 0.20 & 3.39 & 8.33 & \textbf{12.76} \\
        \bottomrule
    \end{tabular}
    }
    \caption{Defense evaluation CIFAR-100 with FPR = $0.005$. OOT stands for out-of-time. The SELENA defense cannot be evaluated on ResNet18 and DenseNet because this defense requires more than half an year to run the experiments. The Mem-guard defense cannot be evaluated against CANARY attack also because of time constraint. MNC stands for Model Not Converging (training failure). DMP defense and Adv-reg defense will sometimes result in extremely low testing accuracy and thus the attack evaluation against these defenses is not meaningful.  Lower PLR while preserving test accuracy is better.}
    
    \label{tab:full_def_eval_tpr_cifar100_fpr0005}
    \vspace{0.2cm}
\end{table*}

\begin{table*}[!t]
    \centering
    \begin{tabular}{lrrrrrrrrrr}
        \toprule
        \multirow{8}{3em}{\centering Dataset-Model} & \multirow{8}{3em}{\centering Defense} & \multirow{8}{3em}{\centering Train Acc.} & \multirow{8}{3em}{\centering Test Acc.}  & \multirow{8}{4em}{ \centering random-perturbation Attack PLR$\downarrow$\\ (FPR@ $0.5\%$) ~\cite{jayaraman2021revisiting}}  & \multirow{8}{3em}{ \centering LOSS Attack PLR$\downarrow$\\ (FPR@ $0.5\%$) ~\cite{yeom2018privacy}} & \multirow{8}{3em}{ \centering modified-entropy Attack PLR$\downarrow$\\ (FPR@ $0.5\%$) ~\cite{song2019membership}}  & \multirow{8}{3em}{ \centering class-NN Attack PLR$\downarrow$\\ (FPR@ $0.5\%$) ~\cite{shokri2017membership}} & \multirow{8}{3em}{\centering LIRA Attack \\ PLR$\downarrow$\\ (FPR@ $0.5\%$)~\cite{carlini2022membership}}  & \multirow{8}{3em}{\centering CANARY Attack PLR$\downarrow$ \\ (FPR@ $0.5\%$)~\cite{wen2022canary}} \\
          & & & & & & \\
          & & & & & & \\
          & & & & & & \\
          & & & & & & \\
          & & & & & & \\
          & & & & & & \\
          & & & & & & \\
         \midrule
         \midrule
        Texas & No-def &  0.6955 & 0.5679 & 0.00 & 0.22 & 1.05 & \textbf{4.99} & 1.21 & N/A \\
        Texas & Adv-reg & 0.3205 & 0.3081 & 0.00 & 0.00 & 0.00 & 4.76 & 1.23 & N/A \\
        Texas & Mem-guard & 0.6955 & 0.5689 & 0.00 & 0.17 & 0.28 & \textbf{4.69} & 1.39 &  N/A \\ 
        Texas & DMP & MNC & MNC & MNC & MNC & MNC & MNC & MNC &  MNC \\
        Texas & MMD & 0.6397 & 0.5598 & 0.00 & 0.23 & 0.23 & \textbf{4.53} & 1.23 & N/A \\ 
        Texas & SELENA & 0.5914 & 0.5366  & 0.00 & 0.10 & 0.15 & \textbf{4.39} & 1.28 & N/A \\ 
        Texas & HAMP & 0.5346 & 0.5190  & 0.00 & 0.25 & 0.35 & \textbf{4.21} & 2.79 & N/A \\
        Texas & DP-SGD & 0.6984 & 0.5682  & 0.00 & 0.03 & 0.03 & \textbf{4.30} & 1.38 & N/A \\
        Texas& MIST & 0.6396 & 0.5638  & 0.00 & 0.23 & 0.39 & \textbf{4.11} & 1.19 & N/A \\ 
        
        \midrule
        Purchase & No-def & 0.9998 & 0.9017 & 0.00 & 0.03 & 0.10 & 1.11 & \textbf{1.31} & N/A \\
        Purchase & Adv-reg & 0.3692 & 0.3636 & 0.00 & 0.93 & 1.01 & \textbf{1.34} & 0.41 &  N/A \\
        Purchase & Mem-guard & 0.9999 & 0.9017 & 0.00 & 0.00 & 0.00 & \textbf{1.38} & 1.32 &  N/A \\
        Purchase & DMP & MNC & MNC & MNC & MNC & MNC & MNC & MNC &  MNC \\
        Purchase & MMD & 0.9897 & 0.9041  & 0.00 & 0.00 & 0.00 & 1.33 & \textbf{1.35} & N/A \\
        Purchase  & SELENA & 0.9234 & 0.8717 & 0.00 & 0.50 & 0.60 & 1.12 & \textbf{1.49} & N/A \\
        Purchase & HAMP & 0.8926 & 0.8353  & 0.00 & 0.99 & 1.01 & \textbf{3.59} & 1.08 & N/A \\
        Purchase & DP-SGD & 0.9893 & 0.8995  & 0.00 & 0.00 & 0.00 & \textbf{1.39} & 1.29 & N/A \\
        Purchase & MIST & 0.9789 & 0.8944  & 0.00 & 0.05 & 0.05 & \textbf{1.05} & 1.19 & N/A \\
        \midrule
        Location & No-def & 1.0000 & 0.7213  & 0.00 & 0.00 & 0.00 & \textbf{1.39} & 1.28 & N/A \\
        Location & Adv-reg & 0.9447 & 0.6410 & 0.00 & 0.06 & 0.13 & \textbf{1.87} & 1.28 &  N/A \\
        Location & Mem-guard & 1.0000 & 0.7143 & 0.00 & 0.00 & 0.00 & \textbf{1.46} & 1.35 &  N/A \\
        Location & DMP & MNC & MNC & MNC & MNC & MNC & MNC & MNC &  MNC \\
        Location & MMD & 1.0000 & 0.6901  & 0.00 & 0.00 & 0.00 & \textbf{1.26} & 0.38 & N/A \\
        Location & SELENA & 0.7307 & 0.6647  & 0.00 & 0.00 & 0.00 & \textbf{3.85} & 2.11 & N/A \\
        Location & HAMP & 0.9885 & 0.7115  & 0.00 & 0.00 & 0.00 & \textbf{1.30} & 1.28 & N/A \\
        Location & DP-SGD & 1.0000 & 0.7113  & 0.00 & 0.00 & 0.00 & \textbf{1.28} & 0.93 & N/A \\
        Location & MIST & 0.9477 & 0.7159  & 0.00 & 0.00 & 0.00 & \textbf{1.11} & 0.93 & N/A \\
        \bottomrule
    \end{tabular}
    \caption{Our proposed defense evaluation on binary feature datasets with FPR = $0.005$. OOT stands for out-of-time. N/A stands for not applicable. The CANARY attack is not applicable to PURCHASE, TEXAS and LOCATION dataset because these three datasets only contain binary features. MNC stands for Model Not Converging (training failure). DMP defense and Adv-reg defense will sometimes result in extremely low testing accuracy and thus the attack evaluation against these defenses is not meaningful. Lower PLR while preserving test accuracy is better.}
    \label{tab:full_def_eval_tpr_binary_fpr0005}
\end{table*}

\begin{figure}
    \centering
    \includegraphics[height=8.5cm,width=0.47\textwidth]{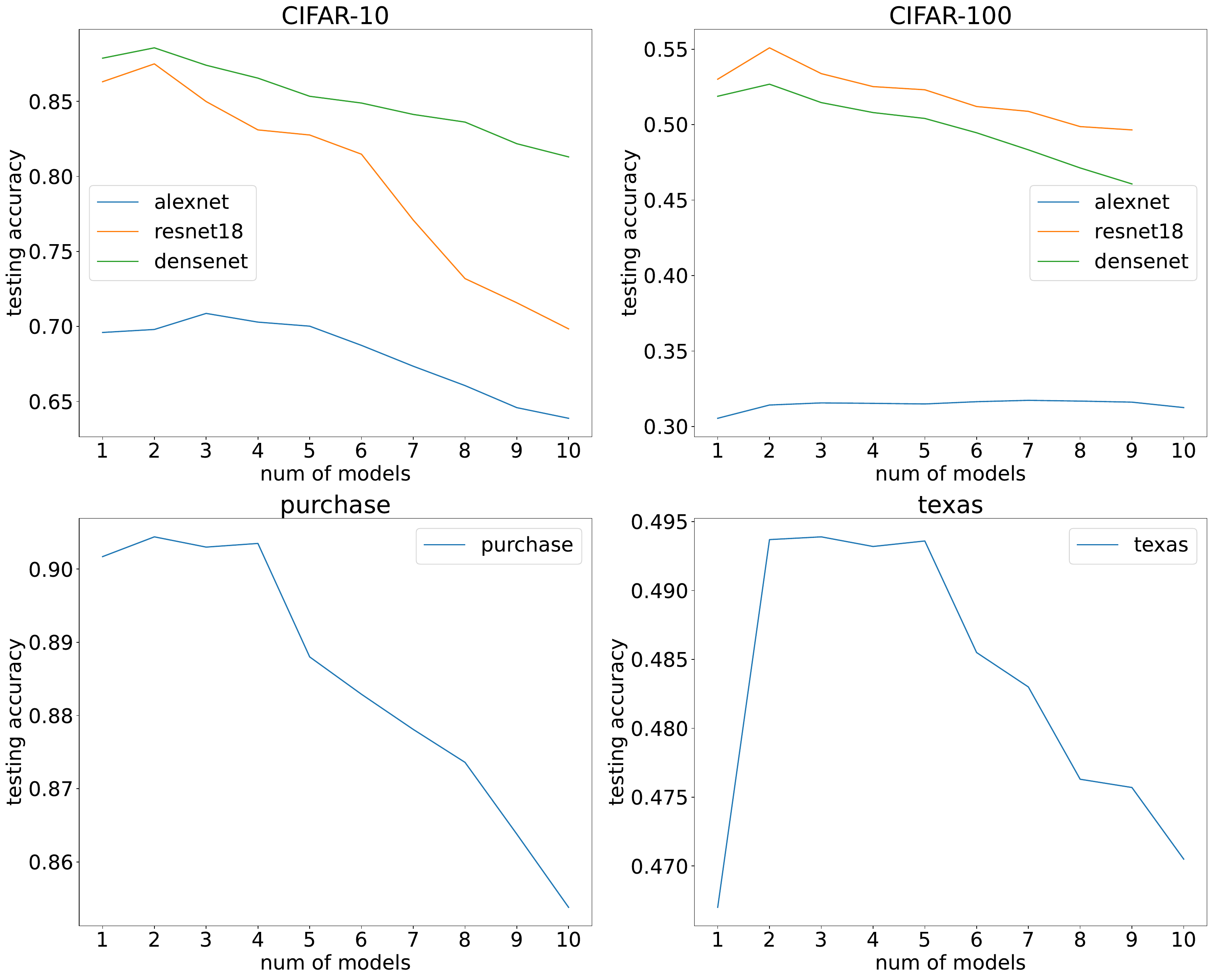}
    \caption{Testing accuracy v.s. number of models for all datasets and models considered in this paper. No mixup data augmentation. The main observation is that testing accuracy can be improved by adding a few models. However, adding too many models would cost testing accuracy.}
    \label{fig:num_user_nomixup}
\end{figure}

\begin{table*}
    \centering
    \normalsize
    \begin{tabular}{lrrrrrrr}
        \toprule
        \multirow{3}{6em}{\centering Dataset-Model} & \multirow{3}{5em}{\centering Number of \\ models} & \multirow{3}{4em}{\centering Train Acc.} & \multirow{3}{4em}{\centering Test Acc.} & \multirow{3}{6em}{ \centering LOSS Attack AUC}
         & \multirow{3}{6em}{\centering LIRA Attack AUC}  & \multirow{3}{7.5em}{\centering LIRA Attack PLR \\ (FPR @ $0.1\%$)} \\
        & & & & & & & \\
        & & & & & & & \\
        \midrule
        CIFAR-10, AlexNet & {1} & 0.9948 & 0.6905 & 0.7213  & 0.7590 & 49.81 \\
        CIFAR-10, AlexNet & {2} & 0.9678 & 0.6931 & 0.6870 & 0.7283 &  43.56 \\
        CIFAR-10, AlexNet & {3} & 0.9631 & 0.6985 & 0.6831 & 0.7237 &  37.88\\
        CIFAR-10, AlexNet & {4} & 0.9620 & 0.6899 & 0.6729  & 0.7218 & 32.01 \\
        CIFAR-10, AlexNet & {5} & 0.9473 & 0.6880 & 0.6590 & 0.7117 & 26.95 \\
        CIFAR-10, AlexNet & {6} & 0.9291 & 0.6873 & 0.6444  & 0.6985  & 22.78 \\
        CIFAR-10, AlexNet & {7} & 0.8991 & 0.6859 & 0.6211  & 0.6826  & 19.59 \\
        CIFAR-10, AlexNet & {8} & 0.8790 & 0.6841 & 0.6113 & 0.6714  &  16.86 \\
        \bottomrule
    \end{tabular}
    \vspace{0.3cm}
    \caption{Experimental results of varying number of models.  CIFAR-10, AlexNet. No cross-difference loss, no mixup data augmentation.}
    \label{tab:num_user}
\end{table*}

\begin{table*}
    \centering
    \normalsize
    \begin{tabular}{lrrrrrrr}
        \toprule
        \multirow{3}{6em}{\centering Dataset-Model} & \multirow{3}{5em}{\centering Number of \\ models} & \multirow{3}{4em}{\centering Train Acc.} & \multirow{3}{4em}{\centering Test Acc.} & \multirow{3}{6em}{ \centering LOSS Attack AUC}
         & \multirow{3}{6em}{\centering LIRA Attack AUC}  & \multirow{3}{7.5em}{\centering LIRA Attack PLR \\ (FPR @ $0.1\%$)} \\
        & & & & & & & \\
        & & & & & & & \\
        \midrule
        CIFAR-100, AlexNet & {1} & 0.9760 & 0.3054 & 0.8899  & 0.9242 & 171.95 \\
        CIFAR-100, AlexNet & {2} & 0.9459 & 0.3142 & 0.8632 & 0.9230 & 170.46 \\
        CIFAR-100, AlexNet & {3} & 0.9105 & 0.3156 & 0.8411 & 0.9114 & 140.80 \\
        CIFAR-100, AlexNet & {4} & 0.9090 & 0.3123 & 0.8334  & 0.9087 & 123.34 \\
        CIFAR-100, AlexNet & {5} & 0.8800 & 0.3149 & 0.8241 & 0.9086  & 121.36\\
        CIFAR-100, AlexNet & {6} & 0.8132 & 0.3164 & 0.7961  & 0.8954 & 95.31 \\
        CIFAR-100, AlexNet & {7} & 0.7314 & 0.3173 & 0.7641  & 0.8729 & 70.19 \\
        CIFAR-100, AlexNet & {8} & 0.6564 & 0.3108 & 0.7268 & 0.8423 & 61.30 \\
        CIFAR-100, AlexNet & {9} & 0.5788 & 0.3161 & 0.6843  & 0.8147 & 55.59 \\
        CIFAR-100, AlexNet & {10} & 0.5223 & 0.3125 & 0.6545  & 0.7874 & 53.28 \\
        CIFAR-100, AlexNet & {11} & 0.4715 & 0.3008 & 0.6315 & 0.7629 & 43.00 \\
        CIFAR-100, AlexNet & {12} & 0.4321 & 0.2920 & 0.6285  & 0.7394 & 36.33 \\
        \bottomrule
    \end{tabular}
    \vspace{0.3cm}
    \caption{Experimental results of varying number of models. CIFAR-100 dataset, AlexNet. No cross-difference loss, no mixup data augmentation.}
    \label{tab:num_user_cifar100}
\end{table*}

\begin{table*}
    \centering
    \normalsize
    \begin{tabular}{lrrrrrrr}
        \toprule
        \multirow{3}{6em}{\centering Dataset-Model} & \multirow{3}{5em}{\centering Number of \\ models} & \multirow{3}{4em}{\centering Train Acc.} & \multirow{3}{4em}{\centering Test Acc.} & \multirow{3}{6em}{ \centering LOSS Attack AUC}
         & \multirow{3}{6em}{\centering LIRA Attack AUC}  & \multirow{3}{7.5em}{\centering LIRA Attack PLR \\ (FPR @ $0.1\%$)} \\
        & & & & & & & \\
        & & & & & & & \\
        \midrule
        CIFAR-10, AlexNet & {1} & 0.9360 & 0.7301 & 0.6535  & 0.7284 & 40.77 \\
        CIFAR-10, AlexNet & {2} & 0.9305 & 0.7235 & 0.6548  & 0.7128 & 33.97  \\
        CIFAR-10, AlexNet & {3} & 0.8980 & 0.7165 & 0.6238  & 0.7076 & 27.01 \\
        CIFAR-10, AlexNet & {4} & 0.8517 & 0.7171 & 0.5935  & 0.6736 & 19.54 \\
        CIFAR-10, AlexNet & {5} & 0.8211 & 0.7057 & 0.5705  & 0.6488 & 17.36 \\
        CIFAR-10, AlexNet & {6} & 0.7940 & 0.6960 & 0.5632  & 0.6284 & 14.61  \\
        CIFAR-10, AlexNet & {7} & 0.7773 & 0.6928 & 0.5583  & 0.6255 & 12.67  \\
        CIFAR-10, AlexNet & {8} & 0.7616 & 0.6853 & 0.5506  & 0.5994 & 9.37  \\
        \bottomrule
    \end{tabular}
    \vspace{0.3cm}
    \caption{Experimental results of varying number of models. Mix-up data augmentation is applied. CIFAR-10, AlexNet.}
    \label{tab:num_user_mixup}
\end{table*}

\begin{table*}
    \centering
    \begin{tabular}{lrrrrrrr}
        \toprule
        \multirow{3}{6em}{\centering Dataset-Model} & \multirow{3}{5em}{\centering Number of \\ models} & \multirow{3}{4em}{\centering Train Acc.} & \multirow{3}{4em}{\centering Test Acc.} & \multirow{3}{6em}{ \centering LOSS Attack AUC}
         & \multirow{3}{6em}{\centering LIRA Attack AUC}  & \multirow{3}{7.5em}{\centering LIRA Attack PLR \\ (FPR @ $0.1\%$)} \\
        & & & & & & & \\
        & & & & & & & \\
        \midrule
        CIFAR-100, AlexNet & {1} & 0.9025 & 0.3680 & 0.8290  & 0.9024  & 256.13 \\
        CIFAR-100, AlexNet & {2} & 0.8466 & 0.3730 & 0.7994  & 0.8815  & 126.38 \\
        CIFAR-100, AlexNet & {3} & 0.7478 & 0.3747 & 0.7376 & 0.8384  & 86.44 \\
        CIFAR-100, AlexNet & {4} & 0.5913 & 0.3610 & 0.6724  & 0.8251 & 77.05\\
        CIFAR-100, AlexNet & {5} & 0.5103 & 0.3418 & 0.6258 & 0.7698 & 46.10\\
        CIFAR-100, AlexNet & {6} & 0.4262 & 0.3201 & 0.5905 & 0.7282 & 29.68 \\
        CIFAR-100, AlexNet & {7} & 0.3878 & 0.3022 & 0.5716 & 0.6951 & 20.60 \\
        CIFAR-100, AlexNet & {8} & 0.3548 & 0.2917 & 0.5616 & 0.6726 & 16.29 \\
        CIFAR-100, AlexNet & {9} & 0.3301 & 0.2733 & 0.5526  & 0.6529 & 12.92 \\
        CIFAR-100, AlexNet & {10} & 0.3083 & 0.2605 & 0.5456  & 0.6391 & 10.93 \\
        \bottomrule
    \end{tabular}
    \vspace{0.3cm}
    \caption{Experimental results of varying number of models. Mix-up data augmentation is applied. CIFAR-100 dataset, AlexNet.}
    \label{tab:num_user_mixup_cifar100}
\end{table*}

\end{document}